\documentclass[12pt]{article}
\usepackage[margin=20truemm]{geometry}
\usepackage[dvipdfmx]{graphicx,color}
\usepackage{bigints}
\usepackage{amsmath,amssymb,bm}
\usepackage{amsthm}
\usepackage{type1cm}
\usepackage{braket}
\usepackage{esvect}
\usepackage{cite}
\usepackage[dvipdfmx]{hyperref}
\usepackage[dvipdfmx]{xcolor}
\hypersetup{%
setpagesize=false,%
bookmarks=true,%
bookmarksdepth=tocdepth,%
bookmarksnumbered=true,%
colorlinks=true,%
linkcolor=black,%
citecolor=black,%
urlcolor=black,%
pdftitle={},%
pdfsubject={},%
pdfauthor={},%
pdfkeywords={}}

\allowdisplaybreaks[3]

\theoremstyle{plain}

\newtheorem*{thm*}{Theorem}
\theoremstyle{definition}

\numberwithin{equation}{section}

\begin{document}

\begin{titlepage}
\begin{normalsize}
\begin{flushright}
\begin{tabular}{l}
KUNS-3002
\end{tabular}
\end{flushright}
  \end{normalsize}
  
\begin{flushright}   \end{flushright} 

~~\\

\vspace*{0cm}
    \begin{Large}
       \begin{center}
       {Gauge invariant discretization of Chern-Simons couplings}
       \end{center}
    \end{Large}
\vspace{1cm}

\begin{center}
	Kohta H{\sc atakeyama},$^{*}$\footnote
           {e-mail address: kohta.hatakeyama@gauge.scphys.kyoto-u.ac.jp}
        Matsuo S{\sc ato}$^{**}$\footnote
           {e-mail address: msato@hirosaki-u.ac.jp} and
	Gota T{\sc anaka},$^{***}$\footnote
           {e-mail address: gotanak@mi.meijigakuin.ac.jp} \\ 
      \vspace{1cm}

         {$^{*}$\it Department of Physics, Kyoto University, Kyoto 606-8502, Japan}\\       
         {$^{**}$\it Graduate School of Science and Technology, Hirosaki University\\ 
 Bunkyo-cho 3, Hirosaki, Aomori 036-8561, Japan}\\
          {$^{***}$\it Institute for Mathematical Informatics, Meiji Gakuin University,\\
1518 Kamikuratacho, Totsuka-ku, Yokohama, Kanagawa 244-8539, Japan}\\
 
 \end{center}

\hspace{5cm}

\begin{abstract}
\noindent
We discretize Chern-Simons couplings in gauge invariant way. We obtain $(p+q)$-forms representing Chern-Simons couplings on  $(p+q)$-simplexes from wedge products of $p$- and $q$-forms on $p$- and $q$-simplexes, respectively, where $p$- and $q$-simplexes form $(p+q)$-simplexes by having a common vertex. We show that the Chern-Simons couplings on simplicial complexes reduce to Chern-Simons couplings on the manifolds in a continuum limit. Moreover, we prove that a typical discretized Chern-Simons term that has the Chern-Simons coupling is gauge invariant.
\end{abstract}

\vfill
\end{titlepage}
\vfil\eject

\setcounter{footnote}{0}

\section{Introduction}\label{intro}
\setcounter{equation}{0}
Superstring theory is a promising candidate for a unified theory including gravity.
One of the recent main problems in string theory
is to determine a true vacuum among extremely large numbers ($> 10^{500}$) of perturbatively stable vacua, which are called the string theory landscape.
If this problem is solved, string theory predicts physical phenomena, which can be captured by experiments and observations.  
One of the most reasonable ways to determine a true vacuum in string theory is to formulate string theory non-perturbatively, derive its effective potential for string backgrounds, and find the minimum. 
Because string backgrounds are the expectation values of the bosonic fields in the supergravities and the D-brane effective theories, the effective potential is a functional of these bosonic fields. Actually, such an effective potential for string backgrounds is derived from string geometry theory, which is one of the candidates of nonpertubative formulations of string theory \cite{Sato:2017qhj, Sato:2019cno, Sato:2020szq, Honda:2020sbl, Honda:2021rcd, Sato:2022owj, Sato:2022brv, Sato:2023lls, Nagasaki:2023fnz}.

The potential energies of manifolds and fields on them are difficult to compare in usual general relativity because they are defined by patching local coordinates. On the other hand,  if we discretize general relativity by Regge calculus \cite{Regge:1961px} or causal dynamical triangulation \cite{Ambjorn:1998xu}, we can globally define manifolds and the fields, and then we can compare the potential energies of them. 

So far, discretization has been performed on the  Einstein-Hilbert action and gauge theories in the standard model, which do not have Chern-Simons couplings. On the other hand, because the supergravities and the D-brane effective theories have Chern-Simons couplings,  we need to discretize Chern-Simons couplings.  Although there are many quantization proposals in \cite{Eliezer:1992sq, Berruto:2000dp, Bietenholz:2002mt, DeMarco:2019pqv, Jacobson:2023cmr} by discretizing the three-dimensional Abelian Chern-Simons theory, which is a special theory where the Chern-Simons coupling becomes topological, our purpose is to just evaluate the values of the effective potentials and actions. Then, it is enough that discretized effective potentials and actions are gauge invariant and reduce to the continuum ones in the continuum limit.  
In this paper, we will discretize Chern-Simons couplings and show that the discretized Chern-Simons couplings are gauge invariant and reduce to continuous Chern-Simons couplings in a continuum limit. This enables us to perform Regge calculus and  causal dynamical triangulation not only on all the bosonic terms in the supergravities and the D-brane effective theories but also on the effective potentials for string backgrounds in superstring theory.

The organization of this paper is as follows. In section 2, we review field strengths of $p$-form fields on simplicial complexes. In section 3, we formulate Chern-Simons couplings on simplicial complexes and show that they reduce to Chern-Simons couplings on manifolds in a continuum limit. In section 4, we prove that a typical discretized Chern-Simons term that has a Chern-Simons coupling defined in the previous section is gauge invariant. In appendixes, we summarize properties and their proofs of the discretized Chern-Simons couplings defined in the main text. 

\section{Field strengths on simplicial complexes}
In this section, we review discretization of the field strength $F_{p+1}$ of a $p$-form field $C_p$ on a simplicial complex. $C_{\mu_1 \mu_2 \dots \mu_p}$ is discretized by a link variable $\Gamma_{C_p}(T^p_{012 \dots p})$ defined on a $p$-simplex $T^p_{012 \dots p}$, where $0, 1, 2, \dots, p$ correspond to vertices $T^0_0, T^0_1, T^0_2, \cdots, T^0_p$. After we define a relation between the continuous variable $C_{\mu_1 \mu_2 \dots \mu_p}$ and the link variable $\Gamma_{C_p}(T^p_{012 \dots p})$, we will show that $F_{\mu_1 \mu_2 \dots \mu_{p+1}}$ is obtained from a plaquette variable on a $(p+1)$-simplex made of $\Gamma_{C_p}(T^p_{012 \dots p})$ in a continuum limit.

The relation between $C_{\mu_1 \mu_2 \dots \mu_p}$ and $\Gamma_{C_p}(T^p_{012 \dots p})$ is defined by 
\begin{align}
	\Gamma_{C_p}(T^p_{012 \dots p})= \exp \left[ i V^{(p)}_{012 \dots p} C^0_{12 \dots p} \left( \frac{a}{p+1}(\vec{v}_1+\vec{v}_2 + \cdots + \vec{v}_p) \right) \right] ,\label{relation}
\end{align}
which corresponds to
$\exp \left[ i \int_{T^p_{012 \dots p}} C_{\mu_1\dots \mu_p} \right]$.
On the $(p+1)$-simplex, we have set 0-coordinates $x^i$ where the origin is $T^0_0$ and the basis vectors are $\vec{v}_i=\frac{1}{a}\overrightarrow{T^0_0 T^0_i} \, (i = 1,2, \dots, p+1)$. The upper index 0 on $C_p$ indicates that $C_p$ is defined on the 0-coordinates.  $a$ is an infinitesimal parameter that gives a continuum limit $a \to 0$. 
$\frac{a}{p+1}(\vec{v}_1+\vec{v}_2 + \cdots + \vec{v}_p)$ is the centroid of $T^p_{012 \dots p}$. $V^{(p)}_{012 \dots p}$ represents the volume of  $T^p_{012 \dots p}$.

The plaquette variable $\Gamma_{F_{p+1}}$ on $T^{p+1}_{012 \dots  p+1}$ corresponding to $F_{p+1}$ is defined by 
\begin{align}
	\Gamma_{F_{p+1}}(T^{p+1}_{012 \dots p+1})
		&\equiv \exp \left[ i \frac{\sqrt{p+2}}{\sqrt{2(p+1)}} \sum_{T^p \subset \partial T^{p+1}_{012 \dots p+1}} \log \Gamma_{C_p}(T^p) \right]
			\nonumber	\\
		&= \exp \left[ i \frac{\sqrt{p+2}}{\sqrt{2(p+1)}} \sum_{q = 0}^{p+1} \frac{(-1)^q}{i}
			\log \Gamma_{C_p}(T^{p,\hat{q}}_{012 \dots p+1}) \right],
		\label{plaquette}
\end{align}
where the orientations of the boundaries are defined by 
\begin{align}
\partial T_{012 \dots p+1}^{p+1} = \sum_{q=0}^{p+1} (-1)^q T^{p, \hat{q} }_{012 \dots p+1},	\label{boundary}
\end{align}
and we have used $\Gamma(-T) = \Gamma^{-1}(T)$. 
$T^{p, \hat{q} }_{012 \dots p+1}$
is a $p$-simplex obtained by omitting a vertex $T^0_q$ from $T_{012 \dots p+1}^{p+1}$. 

From now on, we will show that field strengths are obtained from the plaquette variables in a continuum limit.  
In the following, we will evaluate the exponent of the plaquette up to the leading order in $a$, corresponding to the continuum limit.
First, we consider the $q \neq 0$ case in (\ref{plaquette}), where we set 0-coordinates on the $(p+1)$-simplex.
The link variables in (\ref{plaquette}) are given by 
\begin{align}
	&\Gamma_{C_p}(T^{p, \hat{q}}_{012 \dots p+1})	\nonumber	\\
	=& \exp \left[ i V^{(p)}_{01 \dots {q-1} \hspace{1pt} {q+1} \dots {p+1}} C^0_{1 2 \dots {q-1} \hspace{1pt} {q+1} \dots {p+1}}
		\left( \frac{a}{p+1} (\vec{v}_1 + \vec{v}_2 + \dots + \vec{v}_{q-1} + \vec{v}_{q+1} + \dots + \vec{v}_{p+1}) \right) \right]	\nonumber	\\
	=&\exp \left[ i V^{(p)}_{01 \dots {q-1} \hspace{1pt} {q+1} \dots {p+1}} C^0_{1 2 \dots {q-1} \hspace{1pt} {q+1} \dots {p+1}}
		\left( \vec{n}_{p+1} - \frac{a}{p+1} \vec{v}_q \right) \right]		\nonumber	\\
	=&\exp \left[ i V^{(p)}_{01 \dots {q-1} \hspace{1pt} {q+1} \dots {p+1}} \left(C^0_{1 2 \dots {q-1} \hspace{1pt} {q+1} \dots {p+1}}
		- \frac{a}{p+1} \partial_q 
		C^0_{1 2 \dots {q-1} \hspace{1pt} {q+1} \dots {p+1}} \right)
		\left( \vec{n}_{p+1} \right) \right],	\label{Binx}
\end{align}
where
\begin{align}
	\vec{n}_{p+1} \equiv \frac{a}{p+1} (\vec{v}_1 + \vec{v}_2 + \cdots + \vec{v}_{p+1}).
\end{align}

Next, we consider the $q = 0$ case.  
On the $(p+1)$-simplex, we set 1-coordinates $y^i$ whose origin is $T^0_1$ and which are spanned by
 vectors $\vec{w}_{1}, \vec{w}_{2}, \dots, \vec{w}_{p+1}$ given by
\begin{align}
	\vec{w}_{k} = 
	\begin{cases}
		\frac{1}{a}\overrightarrow{ T^0_1 T^0_k} \ (k \neq 1),	\\
		\vspace{-8pt}	\\
		\frac{1}{a}\overrightarrow{ T^0_1 T^0_0} \ (k = 1).
	\end{cases}
\end{align}
The link variable in (\ref{plaquette}) is given by 
\begin{align}
	\Gamma(T^{p, \hat{0}}_{123 \dots p+1})
	&= \exp \left[ i V^{(p)}_{123 \dots {p+1}} {C}^1_{{2} {3} \dots {p+1}}
		\left( \frac{a }{p+1} (\vec{w}_2 + \vec{w}_3 + \dots + \vec{w}_{p+1}) \right) \right].	\label{Biny}
\end{align}

We transform 1-coordinates to 0-coordinates because we need to calculate in single coordinates to evaluate (\ref{plaquette}). 
A general point $P$ in $T^{p+1}_{012 \dots {p+1} }$ is written as
\begin{align}
	P^0 = x_1 \vec{v}_1 + x_2 \vec{v}_2  \cdots + x_{p+1} \vec{v}_{p+1},	\label{Pinx}
\end{align}
in 0-coordinates and
\begin{align}
	P^1 = y_1 \vec{w}_1 + y_2 \vec{w}_2  \cdots + y_{p+1} \vec{w}_{p+1},	\label{Piny}
\end{align}
in 1-coordinates.
The difference of the origins gives a relation between 
$P^0$ and $P^1$ as
\begin{align}
	P^1 =a\vec{w}_1+P^0.	\label{Piny}
\end{align}
From the relation between $\vec{v}_i$ and $\vec{w}_i$;
\begin{align}
	\vec{v}_1 &= - \vec{w_1}	\\
	\vec{v}_2 &= \vec{w}_2 - \vec{w}_1	\\
	\vec{v}_3 &= \vec{w}_3 - \vec{w}_1	\\
	&\vdots	\nonumber	\\
	\vec{v}_{p+1} &= \vec{w}_{p+1} - \vec{w}_1,
\end{align}
we obtain
\begin{align}
	P^1 &= ( a - x_1 - x_2 - \dots - x_{p+1} ) \vec{w}_{1} + x_2 \vec{w}_2 + x_3 \vec{w}_3 + \cdots + x_{p+1} \vec{w}_{p+1}.	
\end{align} By using \ref{Piny}), we obtain a coordinate transformation, 
\begin{align}
	x_1 &= a - y_1 - y_2 - \cdots - y_{p+1}	\\
	x_n &= y_n \hspace{10pt} (2 \leq n \leq p+1).
\end{align}
By using this, we obtain (\ref{Biny}) in 0-coodinates, 
\begin{align}
&\Gamma(T^{p, \hat{0}}_{123 \dots {p+1}}) \nonumber \\
=
	&\exp \left[ i V^{(p)}_{123 \dots {p+1}} \frac{\partial x^{a_1}}{\partial y^2} \frac{\partial x^{a_2}}{\partial y^3}  \cdots \frac{\partial x^{a_p}}{\partial y^{p+1}} 
		{C}^0_{x_{a_1} x_{a_2} \dots x_{a_p}}
		\left( a \vec{v}_1 + \frac{a}{p+1} (\vec{v}_2 + \vec{v}_3 + \dots + \vec{v}_{p+1} - p \vec{v}_1) \right) \right]	\nonumber	\\
	=&\exp \left[ i V^{(p)}_{123 \dots p+1} \left( C^0_{2 3 \dots {p+1}} - C^0_{1 3 \dots {p+1}} - C^0_{2 1 \dots {p+1}} - \dots - C^0_{2 3 \dots 1} \right)
		\left( \vec{n}_{p+1} \right) \right]	.
\end{align}

To summarize, (\ref{plaquette}) becomes
\begin{align}
	&\Gamma_{F_{p+1}} (T^{p+1}_{012 \dots p+1})	\nonumber	\\
	&= \exp \left[ i \frac{\sqrt{p+2}}{\sqrt{2(p+1)}} V^{(p)}_{123\dots p+1} C^1_{23\dots p+1}(\frac{a}{p+1}(\vec{w}_2 + \vec{w}_3 + \cdots + \vec{w}_{p+1})) \right]
		\nonumber	\\
	&\hspace{10pt} \times \exp \left[ - i \frac{\sqrt{p+2}}{\sqrt{2(p+1)}} V^{(p)}_{023\dots p+1} C^0_{23\dots p+1}(\frac{a}{p+1}(\vec{v}_2 + \vec{v}_3 + \cdots + \vec{v}_{p+1})) \right]	\nonumber	\\
	&\hspace{20pt} \times \exp \left[ i \frac{\sqrt{p+2}}{\sqrt{2(p+1)}} V^{(p)}_{013\dots p+1} C^0_{13\dots p+1}(\frac{a}{p+1}(\vec{v}_1 + \vec{v}_3 + \cdots + \vec{v}_{p+1})) \right]
		\nonumber	\\
	&\hspace{30pt} \times
		\dots \times \exp \left[ (-1)^{p+1} i \frac{\sqrt{p+2}}{\sqrt{2(p+1)}} V^{(p)}_{012\dots p} C^0_{12\dots p}(\frac{a}{p+1}(\vec{v}_1 + \vec{v}_2 + \cdots + \vec{v}_{p})) \right]	\nonumber	\\
	&= \exp \left[ i \frac{\sqrt{p+2}}{\sqrt{2(p+1)}} V^{(p)}_{012\dots p} \biggl( \left\{ C^0_{23\dots p+1} - C^0_{13\dots p+1} - C^0_{21 \dots p+1} - \cdots - C^0_{23 \dots 1} \right\}(\vec{n}_{p+1})
	\right. \biggr. \nonumber\\
		&\hspace{10pt} - \left\{ C^0_{23 \dots p+1} - \frac{a}{p+1}\partial_1 C^0_{23 \dots p+1} \right\}(\vec{n}_{p+1})
			+ \left\{ C^0_{13 \dots p+1} - \frac{a}{p+1}\partial_2 C^0_{13 \dots p+1} \right\}(\vec{n}_{p+1}) + \cdots	\nonumber	\\
		&\hspace{10pt} \left. \left. + (-1)^{p+1} \left\{ C^0_{12 \dots p} - \frac{a}{p+1}\partial_{p+1} C^0_{12 \dots p} \right\}(\vec{n}_{p+1}) \right) \right]
			\nonumber	\\
	&= \exp \left[i \frac{\sqrt{p+2}}{\sqrt{2(p+1)}} \frac{a}{p+1} V^{(p)}_{012\dots p}
		\left\{ \partial_1 C^0_{23\dots p+1} - \partial_2 C^0_{13\dots p+1} + \cdots + (-1)^{p} \partial_{p+1} C^0_{12 \dots p} \right\} (\vec{n}_{p+1})\right]
		\nonumber	\\
	&= \exp \left[i \frac{\sqrt{p+2}}{\sqrt{2(p+1)}} \frac{a}{p+1} V^{(p)}_{012\dots p} F_{12 \dots p+1}(\vec{n}_{p+1}) \right]	\nonumber	\\
	&= \exp \left[i V^{(p+1)}_{012\dots p+1} F_{12 \dots p+1}(\vec{n}_{p+1}) \right].
\end{align}
Note that we have taken a continuum limit where the length of all the edges is $a$ in the second equality and used
\begin{align}
	V^{(p+1)} = \frac{\sqrt{p+2}}{\sqrt{2(p+1)}} \frac{a}{p+1} V^{(p)}
\end{align}
in the last equality.
As a result, we have obtained $F_{p+1}$ from the plaquette variable $\Gamma_{F_{p+1}}$ in the continuum limit $a \to 0$.

\section{Chern-Simons couplings on simplicial complexes}

In this section, we formulate Chern-Simons couplings on simplicial complexes. Explicitly, we display discretization of $H_3 \wedge C_{p-2}$ on a $(p+1)$-simplex $T^{p+1}$. The field strength $H_3$ of a 2-form field $B_2$ is defined on a 3-simplex $T^3$ that is a 3-subsimplex of $T^{p+1}$, whereas a $(p-2)$-form field $C_{p-2}$ is defined on a $(p-2)$-subsimplex $T^{p-2}$. The placement of these subsimplexes is
\begin{align}
	T^3 \cup T^{p-2} &= T^{p+1}	\\
	T^3 \cap T^{p-2} &= T^0_i, \label{share}
\end{align}
analogous to the cup products of cochains\footnote{The authors in \cite{DeMarco:2019pqv, Jacobson:2023cmr} proposed a quantization of the 3-dimensional Chern-Simons theory in a non-compact formulation based on the cup products of cochains.}. 
In this placement, $T^0_i$ is a vertex in $T^{p+1}$ which is necessarily overlapped by the $T^3$ and $T^{p-2}$ because the $p+2$ vertices that consist of the $T^{p+1}$ are less than the $p+3$ vertices that are the 4 vertices that consist of the $T^3$ plus the $p-1$ vertices that consist of the $T^{p-2}$.  

We define a link variable $\Gamma_{H_3 \wedge C_{p-2}}$ on $T^{p+1}$ corresponding to $H_3 \wedge C_{p-2}$ as
\begin{align}
	&\Gamma_{H_3 \wedge C_{p-2}}(T^{p+1}_{012 \dots p+1})	\nonumber	\\
	=& \exp \left[ \frac{i V^{(p+1)}_{012 \dots p+1}}{p+2}
		\sum_{0 \leq \mu_0 < \mu_1 < \mu_2 < \mu_3 \leq p+1} \sum_{0 \leq k \leq 3}
		\sum_{\nu_0 \in \{ \mu_0, \mu_1, \mu_2, \mu_3 \}} \right.
		\nonumber	\\
	&\times \left. \frac{(-1)^{k}}{i V^{(3)}_{\mu_0 \mu_1 \mu_2 \mu_3}} \sqrt{\frac{4}{6}}
			\log \left( \Gamma_{B_2}(T^{2, \hat{\mu}_k}_{\mu_0 \mu_1 \mu_2 \mu_3}) \right)
			\frac{1}{i V^{(p-2)}_{\nu_0 \nu_1 \dots \nu_{p-2}}}
			\log \left( \Gamma_{C_{p-2}} (T^{p-2}_{\nu_0 \nu_1 \dots \nu_{p-2}}) \right) \mathrm{sgn}(\sigma) \right] \label{plaquette2}
\end{align}
where,
\begin{align}
	&0 \leq \nu_1 < \nu_2 < \cdots < \nu_{p-2} \leq p+1,	\\
	&\nu_1, \nu_2, \cdots , \nu_{p-2} \notin \{ \mu_0, \mu_1, \mu_2, \mu_3 \},
\end{align}
and
\begin{align}
	\sigma = \begin{pmatrix}
		\mu_0 & \mu_1 & \mu_2 &\mu_3 & \nu_1 & \nu_2 & \dots & \nu_{p-2}	\\
		0 & 1 & 2 & 3 & 4 & 5& \cdots & p+1	\\
	\end{pmatrix}.
\end{align}
The summation over $k$ implies the summation over the 2-simplexes on the boundary of the 3-simplex, including their orientation.
$T^{2, \hat{\mu}_k}_{\mu_0 \mu_1 \mu_2 \mu_3}$ is a 2-simplex obtained by omitting the vertex $T^0_{\mu_k}$ from the 3-simplex $T^{3}_{\mu_0 \mu_1 \mu_2 \mu_3}$.
The coefficient $1/(p+2)$ comes from the number of choices of a common vertex
between the 3-simplex and the $(p-2)$-simplex where the $H_3$ and $C_{p-2}$ are defined, respectively, among the $p+2$ vertices of the $(p+1)$-simplex. 

In the following, we will see that this link variable gives the wedge product $H_3 \wedge C_{p-2}$ in the continuum limit. As an example, we take a continuum limit when $p=4$. First, on a 5-simplex ${T^5_{012345}}$, we set six coordinates where the vertices $T^0_0,T^0_1,T^0_2,T^0_3,T^0_4,T^0_5$ are the origins,  respectively.  The coordinates with origin $T^0_i$ are spanned by vectors $\vec{v}_{i,1}, \vec{v}_{i,2}, \dots, \vec{v}_{i,5}$ given by
\begin{align}
	\vec{v}_{i,k} =
	\begin{cases}
		\frac{1}{a}\overrightarrow{T^0_{i} T^0_{k}} \ \ (k \neq i),	\\
		\vspace{-8pt}	\\
		\frac{1}{a}\overrightarrow{T^0_{i} T^0_{0}} \ \ (k = i).
	 \end{cases}	
\end{align}

The origin in the coordinates of $T^3$ where $H_3$ is defined is a point whose index is the smallest number, whereas the origin in the coordinates of $T^2$ where  $C_2$ is defined is the common point in (\ref{share}). In these coordinates, (\ref{plaquette2}) becomes 
\begin{align}
	\Gamma_{H_3 \wedge C_2}
	&= \exp \left[ \frac{iV^{(5)}_{012345}}{6} \left\{ H^0_{123}(C^0_{45} + C^1_{45} + C^2_{45} + C^3_{45}) - H^0_{124}(C^0_{35} + C^1_{35} + C^2_{35} + C^4_{35}) \right. \right.
		\nonumber	\\
	&\hspace{30pt} \left. \left. + \cdots + H^2_{345}(C^2_{21} + C^3_{31} + C^4_{41} + C^5_{51}) \right\} \right]. \label{BC2}
\end{align}
Although the arguments of $H_3$ and $C_{p-2}$ are different,  they coincide in the continuum limit because $T^5$ shrinks to a point. Then, we do not describe the arguments. The upper indices of  $H_3$ and $C_{p-2}$  represent the origins in the coordinates. 

If we transform the coordinates to 0-coordinates, (\ref{BC2}) becomes
\begin{align}
	&  \exp \left[ \frac{iV^{(5)}_{012345}}{6} \left\{ H^0_{123}(C^0_{45} + C^0_{45} - C^0_{15} + C^0_{14} + C^0_{45} - C^0_{25} + C^0_{24} + C^0_{45} - C^0_{35} + C^0_{34})  \right. \right. \nonumber	\\
	&\hspace{30pt} - H^0_{124}(C^0_{35} + C^0_{35} - C^0_{15} + C^0_{13} + C^0_{35} - C^0_{25} + C^0_{23} + C^0_{35} - C^0_{45} - C^0_{34}) \nonumber	\\
	&\hspace{30pt} \left. \left. + \cdots + (H^0_{345} - H^0_{245} + H^0_{235} - H^0_{234})( C^0_{12} + C^0_{13} + C^0_{14} + C^0_{15}) \right\} \right]	\nonumber	\\
	&=  \exp \left[ \frac{iV^{(5)}_{012345}}{6} \left\{ 6 H^0_{123}C^0_{45} - 6 H^0_{124}C^0_{35} + \cdots + 6 H^0_{345}C^0_{12}  \right\} \right] \nonumber	\\
	&=  \exp \left[ {iV^{(5)}_{012345}} \left\{ H^0_{123}C^0_{45} - H^0_{124}C^0_{35} + \cdots + H^0_{345}C^0_{12}  \right\} \right].
\end{align}
That is, we have obtained $H_3 \wedge C_{p-2}$ from the link variable in the continuum limit.

\section{Gauge invariance}
In this section, we will prove the gauge invariance of a typical discretized Chern-Simons term that has a discretized Chern-Simons coupling defined in the previous section.

\subsection{$(p+1)$-form field strength}
In this subsection, we will define gauge transformations on the link variables on $p$-simplexes, and prove the gauge invariance of the field strengths of the link variables on $(p+1)$-simplexes.

Gauge transformations on link variables $\Gamma_{C_p}(T^p)$ are defined with gauge parameters $\Gamma_{\Lambda_{p-1}}(T^{p-1})$
by
\begin{align}
	\Gamma'_{C_p}(T^p) = \Gamma_{C_p}(T^p) \prod_{T^{p-1} \subset \partial T^p} \Gamma_{\Lambda_{p-1}}(T^{p-1}).
		\label{gaugeTransformCpLink}
\end{align}
The relation between $\Gamma_{\Lambda_{p-1}}(T^{p-1})$ and the corresponding continuum gauge parameter $\Lambda_{12\dots p-1}$ is 
defined by 
\begin{align}
	\Gamma_{\Lambda_{p-1}}(T^{p-1}_{012\dots p-1}) = \exp \left[ i \sqrt{\frac{p+1}{2p}}
		V^{(p-1)}_{012\dots p-1}\Lambda_{12\dots p-1}^{(0)} \left( \frac{1}{p} (\vec{v}_{0,1} + \vec{v}_{0,2} + \cdots + \vec{v}_{0,p-1}) \right) \right].
\end{align}
As a result, the continuum limit of (\ref{gaugeTransformCpLink})
is given by
\begin{align}
	\exp \left[ i V^{(p)}  C'_p   \right] = \exp \left[ i V^{(p)} \left( C_p +  d\Lambda_{p-1} \right) \right],
\end{align}
where the continuum gauge transformation is realized.

Under (\ref{gaugeTransformCpLink}),
the plaquette variable of the field strength $\Gamma_{F_{p+1}}$ is transformed as 
\begin{align}
	&\Gamma'_{F_{p+1}}(T^{p+1}_{012\dots p+1}) \nonumber		\\
	=& \exp \left[ i \frac{\sqrt{p+2}}{\sqrt{2(p+1)}} \left( \sum_{T^{p}\subset \partial T^{p+1}} \log \Gamma_{C_p}(T^p)
	+\sum_{T^{p}\subset \partial T^{p+1}} \sum_{T^{p-1}\subset \partial T^p}
	\sqrt{\frac{p+1}{2p}} \log \Gamma_{\Lambda_{p-1}}(T^{p-1}) \right) \right].	
\end{align}
Because a $(p-1)$-simplex in a $(p+1)$-complex is boundaries of two $p$-simplexes and the boundaries have opposite orientations,  we have
\begin{align}
	\sum_{T^{p-1}\subset \partial^2 T^{p+1}} \log \Gamma_{\Lambda_{p-1}}(T^{p-1})
	 =& \sum_{r=0}^{p+1} \sum_{q=0, \neq r}^{p+1} \left[ \log \Gamma_{\Lambda_{p-1}}(T^{p-1, \hat{r}, \hat{q}}_{012\dots p+1})
	 	- \log \Gamma_{\Lambda_{p-1}}(T^{p-1, \hat{r}, \hat{q}}_{012\dots p+1}) \right]	\nonumber	\\
	 =& 0,
\end{align}
where we have used (\ref{minus}). Therefore, we have proved the gauge invariance of the plaquette variable of a field strength;
\begin{align}
	\Gamma'_{F_{p+1}}(T^{p+1}_{012\dots p+1}) = \Gamma_{F_{p+1}}(T^{p+1}_{012\dots p+1}).
\end{align}

\subsection{Chern-Simons term}
In this subsection, we will prove the invariance of a typical discretized Chern-Simons term that has a discretized Chern-Simons coupling defined in the previous section under the gauge transformation of the link variable on a $p$-simplex defined in the previous subsection. Explicitly, we will prove the invariance of a discretized Chern-Simons term reducing to $\int_\mathcal{M} C_4 \wedge H_3 \wedge F_3$ in the continuum limit.

In the continuum case, the gauge invariance can be proved as
\begin{align}
	&\delta \int_\mathcal{M} C_4 \wedge H_3 \wedge F_3	\nonumber	\\
	=&\int_\mathcal{M} d\Lambda_3 \wedge H_3 \wedge F_3	\nonumber	\\
	=&\int_\mathcal{M} d(d\Lambda_3 \wedge B_2) \wedge F_3	\nonumber	\\
	=&-\int_\mathcal{M} d\left\{ d(d\Lambda_3 \wedge B_2) \wedge C_2 \right\}	\nonumber	\\
	=& 0.
\end{align}
As in the same steps, it can also be proved in the discretized case as follows.

The discretized Chern-Simons term corresponding to a Chern-Simons term $\int_\mathcal{M} C_4 \wedge H_3 \wedge F_3$ is defined as the Chern-Simons coupling in the previous section,
\begin{align}
	&\sum_{T^{10} \subset \mathcal{M}}\Gamma_{C_4 \wedge H_3 \wedge F_3}(T^{10})	\nonumber	\\
	=&\sum_{T^{10}} \frac{V^{(10)}}{11\cdot8} \sum_{T^4<T^{10}} \sum_{\substack{T^3<T^{10} \\ T^3 \cap T^4 = T^0}}
		\sum_{T^2 \subset \partial T^3} \sum_{\substack{T'^3<T^{10} \\ T'^3 \cap (T^4 \cup T^3) = T'^0}} \sum_{T'^2\subset \partial T'^3}
		\nonumber	\\
	&\times \frac{1}{iV^{(4)}} \log \Gamma_{C_4}(T^4) \frac{1}{iV^{(3)}} \sqrt{\frac{4}{6}} \log \Gamma_{B}(T^2)
		\frac{1}{iV^{(3)}} \sqrt{\frac{4}{6}} \log \Gamma_{C_2}(T'^2) \mathrm{sgn}(\sigma).
		\label{CSterm}
\end{align}
(\ref{CSterm}) is invariant under the transformations of $B_2$ and $C_2$ because $H_3$ and $F_3$ are invariant as in the previous subsection. Under the gauge transformation of $C_4$, (\ref{CSterm}) is transformed as
\begin{align}
&\delta \sum_{T^{10} \subset \mathcal{M}}\Gamma_{C_4 \wedge H_3 \wedge dC_2}(T^{10})
\nonumber \\
=&
	\sum_{T^{10} \subset \mathcal{M}}\Gamma_{d\Lambda_3 \wedge H_3 \wedge F_3}\nonumber \\
	=&\sum_{T^{10}} \frac{V^{(10)}}{11\cdot8} \sum_{T^4<T^{10}} \sum_{T^3\subset \partial T^4} \sum_{\substack{T'^3<T^{10} \\ T'^3 \cap T^4 = T^0}}
		\sum_{T^2 \subset \partial T'^3} \sum_{\substack{T''^3<T^{10} \\ T''^3 \cap (T^4 \cup T'^3) = T'^0}} \sum_{T'^2\subset \partial T''^3}
		\nonumber	\\
	&\times \frac{1}{iV^{(4)}} \log \Gamma_{\Lambda_3}(T^3) \frac{1}{iV^{(3)}} \sqrt{\frac{4}{6}} \log \Gamma_{B_2}(T^2)
		\frac{1}{iV^{(3)}} \sqrt{\frac{4}{6}} \log \Gamma_{C_2}(T'^2) \mathrm{sgn}(\sigma)	\nonumber	\\
	=&\sum_{T^{10}_{012\dots 10}} \frac{V^{(10)}}{11\cdot8} \sum_{0 \leq \mu_0 < \mu_1 < \cdots < \mu_4 \leq 10}
		\sum_{p=0}^{4} \sum_{\substack{\nu_0 \in \{ \mu_i \}}}
		\sum_{\substack{0 \leq \nu_0 < \mu_1 <\nu_2 < \nu_3 \leq 10 \\ \nu_j \notin \{ \mu_i \}}}
		\sum_{q=0}^3 \sum_{\rho_0 \in \{ \mu_i, \nu_j \}}
		\sum_{r=0}^3
		\nonumber	\\
	&\times \frac{(-1)^{p+q+r}}{iV^{(4)}} \log \Gamma_{\Lambda_3}(T^{3, \hat{\mu_p}}_{\mu_0 \mu_1 \dots \mu_4})
		\frac{1}{iV^{(3)}} \sqrt{\frac{4}{6}} \log \Gamma_{B_2}(T^{2, \hat{\nu}_q}_{\nu_0 \nu_1 \nu_2 \nu_3})
		\frac{1}{iV^{(3)}} \sqrt{\frac{4}{6}} \log \Gamma_{C_2}(T^{3, \hat{\rho}_r}_{\rho_0 \rho_1 \rho_2 \rho_3}) \mathrm{sgn}(\sigma)		\nonumber \\
	=& \sum_{T^{10}_{012 \dots 10}} \frac{V^{(10)}}{11} \sum_{0 \leq \mu_0 < \mu_1 < \dots < \mu_7 \leq 10} \frac{1}{iV^{(7)}} \log \Gamma_{d\Lambda_3\wedge H_3}(T^7_{\mu_0 \mu_1 \dots \mu_7})
		\sum_{\rho_0 \in \{\mu_k \}} \sum_{0 \leq r \leq 3} \frac{(-1)^r}{iV^{(3)}} \sqrt{\frac{4}{6}} \log \Gamma_{C_2}(T^{2, \hat{\rho}_r}_{\rho_0 \rho_1 \rho_2 \rho_3}) \mathrm{sgn} (\sigma')
		\nonumber	\\
	=&-\sum_{T^{10} \subset \mathcal{M}} \sqrt{\frac{11}{20}} \sum_{T^9 \subset \partial T^{10}} \frac{1}{i} \log \Gamma_{d(d\Lambda_3 \wedge B_2) \wedge C_2}(T^9),
\label{lastgaugeinv}
\end{align}
where $i = 0,1,2,3,4$, $j =1,2,3$, $k = 0,1,2,3,4,5,6,7$,
\begin{align}
	&0 \leq \rho_1 < \rho_2 < \rho_3 \leq 10, \ \rho_j \notin \{ \mu_k, \nu_j \},	\\
	&\sigma =
		\begin{pmatrix}
		\mu_0 & \mu_1 & \cdots & \mu_4 & \nu_1 & \nu_2 & \nu_3 & \rho_1 & \rho_2 & \rho_3	\\
		0 & 1 & \cdots & \cdots & \cdots & \cdots & \cdots & \cdots & \cdots & 10
		\end{pmatrix},
\end{align}
and
\begin{align}
		&\sigma' =
		\begin{pmatrix}
		\mu_0 & \mu_1 & \cdots & \mu_7 & \rho_1 & \rho_2 & \rho_3	\\
		0 & 1 & \cdots & \cdots & \cdots & \cdots & 10
		\end{pmatrix}.
\end{align}
Link variables $\Gamma_{d\Lambda_3\wedge H_3}(T^7_{012 \dots 7})$ and
$\Gamma_{d(d\Lambda_3 \wedge B_2) \wedge C_2}(T^9)$ are defined as:
\begin{align}
	\Gamma_{d\Lambda_3\wedge H_3}(T^7_{012 \dots 7})
	=& \exp \left[ \frac{iV^{(7)}}{8} \sum_{0 \leq \mu_0 < \mu_1 < \cdots < \mu_4 \leq 7} \sum_{p=0}^4
			\sum_{ \nu_0 \in \{ \mu_i \}} \sum_{q=0}^3 \right.	\nonumber	\\
	&\left. \times		\frac{(-1)^{p+q}}{iV^{(4)}} \log \Gamma_{\Lambda_3}(T^{3, \hat{\mu}_p}_{\mu_0\mu_1\mu_2\mu_3\mu_4})
			\frac{1}{iV^{(3)}} \sqrt{\frac{4}{6}} \log \Gamma_{B_2}(T^{2, \hat{\nu}_q}_{\nu_0\nu_1 \nu_2 \nu_3})
			\mathrm{sgn}(\sigma'') \right],
\end{align}
where $0 \leq \nu_1 < \nu_2 < \nu_3 \leq 7, \ \nu_j \notin \{ \mu_i \}$, and
\begin{align}
	&\sigma'' =
		\begin{pmatrix}
		\mu_0 & \mu_1 & \cdots & \mu_4 & \nu_1 & \nu_2 & \nu_3	\\
		0 & 1 & \cdots & \cdots & \cdots & \cdots & 7
		\end{pmatrix},
\end{align}
and
\begin{align}
	\Gamma_{d(d\Lambda_3 \wedge B_2) \wedge C_2}(T^9_{012 \dots 9})
		=&\exp \left[ i \frac{V^{(9)}}{10} \sum_{0 \leq \mu_0 < \mu_1 < \cdots < \mu_7 \leq 9}
			\sum_{p = 0}^7
			\sum_{\nu_0 \in \{ \mu_k \}} \right.	\nonumber	\\
		&\left. \times \frac{(-1)^p}{iV^{(7)}} \sqrt{\frac{8}{14}}
			\log \Gamma_{d\Lambda_3 \wedge B_2}(T^{6, \hat{\mu}_p}_{\mu_0 \mu_1 \dots \mu_7})
			\frac{1}{iV^{(2)}} \log \Gamma_{C_2}(T^2_{\nu_0 \nu_1 \nu_2})
			\mathrm{sgn}(\sigma''') \right],
			\label{2nd}
\end{align}
where $0 \leq \nu_1 < \nu_2 \leq 9, \ \nu_1, \nu_2 \notin \{ \mu_k \}, $ and
\begin{align}
	\sigma''' =
		\begin{pmatrix}
		\mu_0 & \mu_1 & \cdots & \mu_7 & \nu_1 & \nu_2	\\
		0 & 1 & \cdots & \cdots & \cdots & 7
		\end{pmatrix}.
\end{align}
In the last equality in \eqref{lastgaugeinv}, we have used (\ref{dLambdaWedgeB}).
Here, (\ref{minus}) and (\ref{diff}) lead
$\Gamma_{d\Lambda_3 \wedge B}(T^6)
=\Gamma^{-1}_{d\Lambda_3 \wedge B}(-T^6) $.
Then, (\ref{2nd}) implies
\begin{equation}
\Gamma_{d(d\Lambda_3 \wedge B_2) \wedge C_2}(T^9) = \Gamma^{-1}_{d(d\Lambda_3 \wedge B_2) \wedge C_2}(-T^9).
\end{equation}
Then, (\ref{lastgaugeinv})=0
because $T^8$ in  $T^{10}$ is boundaries of two $T^9$s and the boundaries have opposite orientations.
Therefore, we have proved the gauge invariance of the Chern-Simons term (\ref{CSterm}).

\section{Conclusion}
We have discretized Chern-Simons couplings in a gauge invariant way. We have obtained $(p+q)$-forms representing Chern-Simons couplings on $(p+q)$-simplexes from wedge products of $p$- and $q$-forms on $p$- and $q$-simplexes, respectively, where $p$- and $q$-simplexes form $(p+q)$-simplexes by having a common point. We have shown that the Chern-Simons couplings on simplicial complexes reduce to Chern-Simons couplings on the manifolds in a continuum limit.   Moreover, we have proven that a typical discretized Chern-Simons term that has the Chern-Simons coupling is gauge invariant. We have also proven some properties of the Chern-Simons couplings.

Based on these results, we can perform Regge calculus on the effective potentials for string backgrounds and find their minimums by numerical simulations \cite{future}. As a result, we may determine a true vacuum in string theory.

\section{Acknowledgement}
We would like to thank
I. Kanamori,
H. Kawai,
T. Masuda,
K. Nagasaki,
J. Nishimura,
Y. Sugimoto,
M. Takeuchi,
T. Yoneya,
and especially
A. Tsuchiya
for discussions.
The work of M. S. is supported by Hirosaki University Priority Research Grant for Future Innovation. The work of G.T. is supported by JSPS KAKENHI Grant Numbers JP22H01222.

\appendix

\section{Wedge product on simplex}
A wedge product $A \wedge B$ of $p$-form $A$ and $q$-form $B$ is discretized on the $(p+q)$-simplex $T^{p+q}_{012\dots p+q}$ by
a link variable,
\begin{align}
	&\Gamma_{A \wedge B}(T^{p+q}_{012\dots p+q})	\nonumber	\\
	\equiv& \exp \left[ \sum_{T^p < T^{p+q}} \sum_{\substack{T^q < T^{p+q} \\ T^q \cap T^p = T^0}}
		\frac{V^{(p+q)}}{p+q+1} \frac{1}{iV^{(p)}} \log \Gamma_A(T^p)
		\frac{1}{iV^{(q)}} \log \Gamma_B(T^q) \mathrm{sgn}(\sigma) \right]	\label{LinkAB}\\
	=&\exp \left[ \sum_{0 \leq \mu_0 < \mu_1 < \dots < \mu_p \leq p+q} \sum_{\nu_0 \in \{ \mu_0, \mu_1, \dots, \mu_{p} \}} \right.
		\nonumber	\\
	&\left. \times \frac{V^{(p+q)}}{p+q+1} \frac{1}{iV^{(p)}} \log \Gamma_A(T^p_{\mu_0\mu_1 \dots \mu_p})
		\frac{1}{iV^{(q)}} \log \Gamma_B(T^q_{\nu_0 \nu_1 \dots \nu_q}) \mathrm{sgn}(\sigma) \right] ,	\label{LinkAB_vertex}
		\end{align}
where
$0 \leq \nu_1 < \nu_2 < \dots < \nu_q \leq p+q$,
$\{ \nu_1, \nu_2, \dots, \nu_q \} = \{0,1,2,\dots,p+q \} / \{ \mu_0, \mu_1, \dots, \mu_p \}$, and
\begin{align}
	&\sigma =
		\begin{pmatrix}
		\mu_0 & \mu_1 & \cdots & \mu_{p} & \nu_1 & \nu_2 & \cdots & \nu_q	\\
		0 & 1 & \cdots & \cdots & \cdots & \cdots & \cdots & p+q
		\end{pmatrix}. \nonumber
\end{align}
$\mu_0, \mu_1, \dots, \mu_p$ are vertices of $p$-simplex $T^p<T^{p+q}$ where $A$ is defined,
$\nu_0, \nu_1, \dots, \nu_q$ are vertices of $q$-simplex $T^q<T^{p+q}$ where $B$ is defined, and $\nu_0$ is also shared by $T^p$.
Note that $\nu_1, \nu_2, \dots, \nu_q $ are automatically fixed once we determine $\mu_0, \mu_1, \dots \mu_p$
because $\{ \nu_1, \nu_2, \dots, \nu_q \}$ is the complement of $\{\mu_0, \mu_1, \dots \mu_p\}$.
The summations over $\mu_0, \mu_1, \dots \mu_p$ and $\nu_0$ in \eqref{LinkAB_vertex} is
equivalent to those over $T^p$ and $T^q$
in \eqref{LinkAB}.

In the continuum case, let $A$ be a $p$-form and $B$ be a $q$-form, then a wedge product $A \wedge B$ satisfies
\begin{align}
	A \wedge B = (-1)^{pq} B \wedge A.	\label{CommutativityofWedgeProduct}
\end{align}
We will show the analogous fact that the link variable $\Gamma_{A\wedge B}$ of the wedge product $A \wedge B$ satisfies
\begin{align}
\Gamma_{A \wedge B} = \Gamma^{(-1)^{pq}}_{B \wedge A}.	\label{CommutativityofWedgeLink}
\end{align}
We begin with rewriting the summations over vertices in \eqref{LinkAB_vertex} as
\begin{align}
	&\Gamma_{A \wedge B}(T^{p+q}_{012\dots p+q})	\nonumber	\\
	=&\exp \left[ \sum_{0 \leq \mu_0 < \mu_1 < \dots < \mu_p \leq p+q} \sum_{\nu_0 \in \{ \mu_0, \mu_1, \dots, \mu_{p} \}} \right.
		\nonumber	\\
	&\left. \times \frac{V^{(p+q)}}{p+q+1} \frac{1}{iV^{(p)}} \log \Gamma_A(T^p_{\mu_0\mu_1 \dots \mu_p})
		\frac{1}{iV^{(q)}} \log \Gamma_B(T^q_{\nu_0 \nu_1 \dots \nu_q}) \mathrm{sgn}(\sigma) \right]	\nonumber	\\
	=&\exp \left[ \sum_{\nu_0 \in \{ 0,1,2,\dots p+q \}}
		\sum_{\substack{0 \leq \mu_0 < \mu_1 < \dots < \mu_{p-1} \leq p+q}} \right.
		\nonumber	\\
	&\left. \times \frac{V^{(p+q)}}{p+q+1} \frac{1}{iV^{(p)}} \log \Gamma_A(T^p_{\mu_0\mu_1 \dots \mu_k \nu_0 \mu_{k+1} \cdots \mu_{p-1}})
		\frac{1}{iV^{(q)}} \log \Gamma_B(T^q_{\nu_0 \nu_1 \dots \nu_q}) \mathrm{sgn}(\sigma') \right],
\end{align}
where
\begin{gather}
	\mu_k < \nu_0 < \mu_{k+1}, \ \nu_l < \nu_0 < \nu_{l+1},	\\
	0 \leq \nu_1 < \nu_2 < \dots  < \nu_q \leq p+q,\\
	\{ \nu_1, \nu_2 , \dots , \nu_q \} = \{ 0,1,2, \dots , p+q \} / \{ \nu_0, \mu_0, \mu_1, \dots, \mu_{p-1} \},
\end{gather}
and
\begin{gather}
	\sigma' =
		\begin{pmatrix}
		\mu_0 & \mu_1 & \cdots & \mu_k & \nu_0 & \mu_{k+1} & \cdots & \mu_{p-1} & \nu_1 & \nu_2 & \cdots & \nu_q	\\
		0 & 1 & \cdots & \cdots & \cdots & \cdots & \cdots & \cdots & \cdots & \cdots & \cdots & p+q
		\end{pmatrix}.
\end{gather}
Due to the antisymmetry $\Gamma_A(T^p) = \Gamma^{-1}_A(-T^p)$ and $\Gamma_B(T^q) = \Gamma^{-1}_B(-T^q)$, we have
\begin{align}
	\log \Gamma_A(T^p_{\mu_0\mu_1 \dots \mu_k \nu_0 \mu_{k+1} \cdots \mu_{p-1}})
	=&(-1)^{k+1} \log \Gamma_A(T^p_{\nu_0 \mu_0\mu_1 \cdots \mu_{p-1}}),	\\
	\log \Gamma_B(T^q_{\nu_0 \nu_1 \dots \nu_q})
	=& (-1)^{l} \log \Gamma_B(T^q_{ \nu_1 \dots \nu_l \nu_0 \nu_{l+1} \cdots \nu_q}).
\end{align}
Moreover, we can rewrite $\sigma$ as
\begin{align}
	&\sigma' =
		\begin{pmatrix}
		\mu_0 & \mu_1 & \cdots & \mu_k & \nu_0 & \mu_{k+1} & \cdots & \mu_{p-1} & \nu_1 & \nu_2 & \cdots & \nu_q	\\
		0 & 1 & \cdots & \cdots & \cdots & \cdots & \cdots & \cdots & \cdots & \cdots & \cdots & p+q
		\end{pmatrix}	\nonumber	\\
		=& (-1)^{k+1}
		\begin{pmatrix}
		\nu_0 &\mu_0 & \mu_1 & \cdots & \mu_{p-1} & \nu_1 & \nu_2 & \cdots & \nu_q	\\
		0 & 1 & \cdots & \cdots & \cdots & \cdots & \cdots & \cdots & p+q
		\end{pmatrix}	\nonumber	\\
		=& (-1)^{(k+1) + pq}
		\begin{pmatrix}
		 \nu_0 &  \nu_1 & \nu_2 & \cdots & \nu_q &\mu_0 & \mu_1 & \cdots & \mu_{p-1}	\\
		0 & 1 & \cdots & \cdots & \cdots & \cdots & \cdots & \cdots & p+q
		\end{pmatrix}	\nonumber	\\
		=& (-1)^{(k+1) + l + pq}
		\begin{pmatrix}
		 \nu_1 & \nu_2 & \cdots & \nu_l & \nu_0 & \nu_{l+1} & \cdots & \nu_q &\mu_0 & \mu_1 & \cdots & \mu_{p-1}	\\
		0 & 1 & \cdots & \cdots & \cdots & \cdots & \cdots & \cdots & \cdots & \cdots & \cdots & p+q
		\end{pmatrix}	\nonumber	\\
		\equiv& (-1)^{(k+1) + l + pq} \sigma''.
\end{align}
Using these results, we obtain
\begin{align}
	\Gamma_{A\wedge B}(T^{p+q})
	=&\exp \left[ \sum_{\nu_0 \in \{ 0,1,2,\dots p+q \}}
		\sum_{\substack{0 \leq \mu_0 < \mu_1 < \dots < \mu_{p-1} \leq p+q}}
		\frac{V^{(p+q)}}{p+q+1} \frac{1}{iV^{(p)}} (-1)^{k+1} \log \Gamma_A(T^p_{\nu_0 \mu_0\mu_1 \cdots \mu_{p-1}}) \right.
		\nonumber	\\
	&\left. \times
		\frac{1}{iV^{(q)}}
		(-1)^{l} \log \Gamma_B(T^q_{ \nu_1 \dots \nu_l \nu_0 \nu_{l+1} \cdots \nu_q})
		(-1)^{(k+1) + l + pq}
		\mathrm{sgn}(\sigma'') \right]	\nonumber	\\
	=&\exp \left[ \sum_{\nu_0 \in \{ 0,1,2,\dots p+q \}}
		\sum_{\substack{0 \leq \nu_1 < \nu_2 < \dots < \nu_{q} \leq p+q}}
		\frac{V^{(p+q)}}{p+q+1} \frac{1}{iV^{(p)}} (-1)^{pq} \log \Gamma_A(T^p_{\nu_0 \mu_0\mu_1 \cdots \mu_{p-1}}) \right.
		\nonumber	\\
	&\left. \times 
		\frac{1}{iV^{(q)}} 
		\log \Gamma_B(T^q_{ \nu_1 \dots \nu_l \nu_0 \nu_{l+1} \cdots \nu_q})
		\mathrm{sgn}(\sigma'') \right]	\nonumber	\\
	=&\exp \left[ \sum_{\substack{0 \leq \nu_0 < \nu_1 < \dots < \nu_q \leq p+q}}
		\sum_{\mu \in \{ \nu_0, \nu_1, \dots, \nu_q \}}
		\frac{V^{(p+q)}}{p+q+1} \frac{1}{iV^{(p)}} (-1)^{pq} \log \Gamma_A(T^p_{\mu \mu_0\mu_1 \cdots \mu_{p-1}}) \right.
		\nonumber	\\
	&\left. \times
		\frac{1}{iV^{(q)}}
		\log \Gamma_B(T^q_{ \nu_0 \nu_1 \dots \nu_q})
		\mathrm{sgn}(\sigma'') \right]	\nonumber	\\
	=& \Gamma_{B \wedge A}^{(-1)^{pq}}(T^{p+q}).
\end{align}

\section{$\Gamma_{A\wedge B}(-T^{p+q}) = \Gamma^{-1}_{A \wedge B}(T^{p+q})$}
To show that the link variable $\Gamma_{A\wedge B}(T^{p+q})$ of $p$-form $A$ and $q$-form $B$ satisfies
\begin{align}
	\Gamma_{A\wedge B}(-T^{p+q}) = \Gamma^{-1}_{A \wedge B}(T^{p+q}), \label{minus}
\end{align}
we will prove
\begin{align}
	\Gamma^{-1}_{A \wedge B}(T^{p+q}_{012\dots p+q}) = \left. \Gamma_{A \wedge B}(T^{p+q}_{012\dots p+q}) \right|_{\substack{0 \leftrightarrow 1}},
\end{align}
since $T^{p+q}_{012\dots p+q} = - T^{p+q}_{102\dots p+q}$.

First, we decompose the summation over $\mu_0, \mu_1, \dots \mu_p$ and $\nu_0$ in \eqref{LinkAB_vertex} into the sum of the following eight parts
$P_1, P_2, \dots, P_8$ given in the following:
\begin{equation}
\log(\Gamma^{-1}_{A\wedge B}(T^{p+q}_{012\dots p+q}))=P_1 + P_2 + \cdots + P_8,
\end{equation}
and then exchange the labels of vertices as $0 \leftrightarrow 1$.
\begin{enumerate}
	\item $P_1$ is a part of the summation where $\mu_0 = 0, \ \mu_1 = 1$, and $\nu_0 = 0$ and given by
		\begin{align}
			P_1=&\sum_{2 \leq \mu_2 < \mu_3 < \dots < \mu_p \leq p+q}
				\frac{V^{(p+q)}}{p+q+1} \frac{1}{iV^{(p)}} \log \Gamma_A(T^p_{01 \mu_2 \mu_3 \dots \mu_p})
			\frac{1}{iV^{(q)}} \log \Gamma_B(T^q_{0 \nu_1 \dots \nu_q})	\nonumber	\\
			&\times \mathrm{sgn}
				\begin{pmatrix}
					0 & 1 & \mu_2 & \mu_3 & \dots & \mu_p & \nu_1 & \nu_2 & \dots & \nu_q	\\
					0 & 1 & 2 & 3 & \dots & \dots & \dots & \dots & \dots & p+q
				\end{pmatrix}.
		\end{align}
		\item $P_2$ is a part of the summation where $\mu_0 = 0, \ \mu_1 = 1$, and $\nu_0 = 1$ and given by
			\begin{align}
			P_2=&\sum_{2 \leq \mu_2 < \mu_3 < \dots < \mu_p \leq p+q}
				\frac{V^{(p+q)}}{p+q+1} \frac{1}{iV^{(p)}} \log \Gamma_A(T^p_{01 \mu_2 \mu_3 \dots \mu_p})
			\frac{1}{iV^{(q)}} \log \Gamma_B(T^q_{1 \nu_1 \dots \nu_q})	\nonumber	\\
			&\times \mathrm{sgn}
				\begin{pmatrix}
					0 & 1 & \mu_2 & \mu_3 & \dots & \mu_p & \nu_1 & \nu_2 & \dots & \nu_q	\\
					0 & 1 & 2 & 3 & \dots & \dots & \dots & \dots & \dots & p+q
				\end{pmatrix}.	\nonumber	\\
				\mbox{Then, }
			P_1|_{0\leftrightarrow 1}= &\sum_{2 \leq \mu_2 < \mu_3 < \dots < \mu_p \leq p+q}
				\frac{V^{(p+q)}}{p+q+1} \frac{1}{iV^{(p)}} \log \Gamma_A(T^p_{10 \mu_2 \mu_3 \dots \mu_p})
			\frac{1}{iV^{(q)}} \log \Gamma_B(T^q_{1 \nu_1 \dots \nu_q})	\nonumber	\\
			&\times \mathrm{sgn}
				\begin{pmatrix}
					1 & 0 & \mu_2 & \mu_3 & \dots & \mu_p & \nu_1 & \nu_2 & \dots & \nu_q	\\
					1 & 0 & 2 & 3 & \dots & \dots & \dots & \dots & \dots & p+q
				\end{pmatrix}	\nonumber	\\
			=& -\sum_{2 \leq \mu_2 < \mu_3 < \dots < \mu_p \leq p+q}
				\frac{V^{(p+q)}}{p+q+1} \frac{1}{iV^{(p)}} \log \Gamma_A(T^p_{01 \mu_2 \mu_3 \dots \mu_p})
			\frac{1}{iV^{(q)}} \log \Gamma_B(T^q_{1 \nu_1 \dots \nu_q})	\nonumber	\\
			&\times \mathrm{sgn}
				\begin{pmatrix}
					0 & 1 & \mu_2 & \mu_3 & \dots & \mu_p & \nu_1 & \nu_2 & \dots & \nu_q	\\
					0 & 1 & 2 & 3 & \dots & \dots & \dots & \dots & \dots & p+q
				\end{pmatrix}	\nonumber	\\
			=&-P_2, \nonumber \\
	\mbox{and }		P_2|_{0\leftrightarrow 1} =&\sum_{2 \leq \mu_2 < \mu_3 < \dots < \mu_p \leq p+q}
				\frac{V^{(p+q)}}{p+q+1} \frac{1}{iV^{(p)}} \log \Gamma_A(T^p_{10 \mu_2 \mu_3 \dots \mu_p})
			\frac{1}{iV^{(q)}} \log \Gamma_B(T^q_{0 \nu_1 \dots \nu_q})	\nonumber	\\
			&\times \mathrm{sgn}
				\begin{pmatrix}
					1 & 0 & \mu_2 & \mu_3 & \dots & \mu_p & \nu_1 & \nu_2 & \dots & \nu_q	\\
					1 & 0 & 2 & 3 & \dots & \dots & \dots & \dots & \dots & p+q
				\end{pmatrix}	\nonumber	\\
			=& -\sum_{2 \leq \mu_2 < \mu_3 < \dots < \mu_p \leq p+q}
				\frac{V^{(p+q)}}{p+q+1} \frac{1}{iV^{(p)}} \log \Gamma_A(T^p_{01 \mu_2 \mu_3 \dots \mu_p})
			\frac{1}{iV^{(q)}} \log \Gamma_B(T^q_{0 \nu_1 \dots \nu_q})	\nonumber	\\
			&\times \mathrm{sgn}
				\begin{pmatrix}
					0 & 1 & \mu_2 & \mu_3 & \dots & \mu_p & \nu_1 & \nu_2 & \dots & \nu_q	\\
					0 & 1 & 2 & 3 & \dots & \dots & \dots & \dots & \dots & p+q
				\end{pmatrix}	\nonumber	\\
			=&-P_1.
			\end{align}
		\item $P_3$ is a part of the summation where $\mu_0 = 0, \ \mu_1 = 1$, and $\nu_0 \neq 0, 1$, and given by
			\begin{align}
			P_3=&\sum_{2 \leq \mu_2 < \mu_3 < \dots < \mu_p \leq p+q} \sum_{\nu_0 \in \{ \mu_2, \mu_3, \dots , \mu_p \}}
				\frac{V^{(p+q)}}{p+q+1} \frac{1}{iV^{(p)}} \log \Gamma_A(T^p_{01 \mu_2 \mu_3 \dots \mu_p})
			\frac{1}{iV^{(q)}} \log \Gamma_B(T^q_{\nu_0 \nu_1 \dots \nu_q})	\nonumber	\\
			&\times \mathrm{sgn}
				\begin{pmatrix}
					0 & 1 & \mu_2 & \mu_3 & \dots & \mu_p & \nu_1 & \nu_2 & \dots & \nu_q	\\
					0 & 1 & 2 & 3 & \dots & \dots & \dots & \dots & \dots & p+q
				\end{pmatrix}.	\nonumber	\\
		\mbox{Then, }	P_3|_{0\leftrightarrow 1} =&\sum_{2 \leq \mu_2 < \mu_3 < \dots < \mu_p \leq p+q}
				\frac{V^{(p+q)}}{p+q+1} \frac{1}{iV^{(p)}} \log \Gamma_A(T^p_{10 \mu_2 \mu_3 \dots \mu_p})
			\frac{1}{iV^{(q)}} \log \Gamma_B(T^q_{\nu_0 \nu_1 \dots \nu_q})	\nonumber	\\
			&\times \mathrm{sgn}
				\begin{pmatrix}
					1 & 0 & \mu_2 & \mu_3 & \dots & \mu_p & \nu_1 & \nu_2 & \dots & \nu_q	\\
					1 & 0 & 2 & 3 & \dots & \dots & \dots & \dots & \dots & p+q
				\end{pmatrix}	\nonumber	\\
			=& -\sum_{2 \leq \mu_2 < \mu_3 < \dots < \mu_p \leq p+q}
				\frac{V^{(p+q)}}{p+q+1} \frac{1}{iV^{(p)}} \log \Gamma_A(T^p_{01 \mu_2 \mu_3 \dots \mu_p})
			\frac{1}{iV^{(q)}} \log \Gamma_B(T^q_{\nu_0 \nu_1 \dots \nu_q})	\nonumber	\\
			&\times \mathrm{sgn}
				\begin{pmatrix}
					0 & 1 & \mu_2 & \mu_3 & \dots & \mu_p & \nu_1 & \nu_2 & \dots & \nu_q	\\
					0 & 1 & 2 & 3 & \dots & \dots & \dots & \dots & \dots & p+q
				\end{pmatrix}	\nonumber	\\
			=&-P_3
			\end{align}
	\item $P_4$ is a part of the summation where $\mu_0 = 0, \ \nu_1 = 1$, and $\nu_0 = 0$, and given by 
		\begin{align}
			P_4=&\sum_{2 \leq \mu_1 < \mu_2 < \dots < \mu_p \leq p+q}
				\frac{V^{(p+q)}}{p+q+1} \frac{1}{iV^{(p)}} \log \Gamma_A(T^p_{0 \mu_1 \mu_2 \dots \mu_p})
			\frac{1}{iV^{(q)}} \log \Gamma_B(T^q_{0 1 \dots \nu_q})	\nonumber	\\
			&\times \mathrm{sgn}
				\begin{pmatrix}
					0 & \mu_1 & \mu_2 & \mu_3 & \dots & \mu_p & 1 & \nu_2 & \dots & \nu_q	\\
					0 & 1 & 2 & 3 & \dots & \dots & \dots & \dots & \dots & p+q
				\end{pmatrix}.
		\end{align}
			\item $P_5$  is a part of the summation where $\mu_0 = 1, \ \nu_1 = 0$, and $\nu_0 = 1$, and given by
		\begin{align}
			P_5=&\sum_{2 \leq \mu_1 < \mu_2 < \dots < \mu_p \leq p+q}
				\frac{V^{(p+q)}}{p+q+1} \frac{1}{iV^{(p)}} \log \Gamma_A(T^p_{1 \mu_1 \mu_2 \dots \mu_p})
			\frac{1}{iV^{(q)}} \log \Gamma_B(T^q_{10 \dots \nu_q})	\nonumber	\\
			&\times \mathrm{sgn}
				\begin{pmatrix}
					1 & \mu_1 & \mu_2 & \mu_3 & \dots & \mu_p & 0 & \nu_2 & \dots & \nu_q	\\
					0 & 1 & 2 & 3 & \dots & \dots & \dots & \dots & \dots & p+q
				\end{pmatrix}.	\nonumber	\\
	\mbox{Then, }			P_4|_{0\leftrightarrow 1} =&\sum_{2 \leq \mu_1 < \mu_2 < \dots < \mu_p \leq p+q}
				\frac{V^{(p+q)}}{p+q+1} \frac{1}{iV^{(p)}} \log \Gamma_A(T^p_{1 \mu_1 \mu_2 \dots \mu_p})
			\frac{1}{iV^{(q)}} \log \Gamma_B(T^q_{1 0 \dots \nu_q})	\nonumber	\\
			&\times \mathrm{sgn}
				\begin{pmatrix}
					1 & \mu_1 & \mu_2 & \mu_3 & \dots & \mu_p & 0 & \nu_2 & \dots & \nu_q	\\
					1 & 0 & 2 & 3 & \dots & \dots & \dots & \dots & \dots & p+q
				\end{pmatrix}	\nonumber	\\
			=&-\sum_{2 \leq \mu_1 < \mu_2 < \dots < \mu_p \leq p+q}
				\frac{V^{(p+q)}}{p+q+1} \frac{1}{iV^{(p)}} \log \Gamma_A(T^p_{1 \mu_1 \mu_2 \dots \mu_p})
			\frac{1}{iV^{(q)}} \log \Gamma_B(T^q_{1 0 \dots \nu_q})	\nonumber	\\
			&\times \mathrm{sgn}
				\begin{pmatrix}
					1 & \mu_1 & \mu_2 & \mu_3 & \dots & \mu_p & 0 & \nu_2 & \dots & \nu_q	\\
					0 & 1 & 2 & 3 & \dots & \dots & \dots & \dots & \dots & p+q
				\end{pmatrix}	\nonumber	\\
			=&- P_5, \nonumber \\
	\mbox{and }		P_5|_{0\leftrightarrow 1} =&\sum_{2 \leq \mu_1 < \mu_2 < \dots < \mu_p \leq p+q}
				\frac{V^{(p+q)}}{p+q+1} \frac{1}{iV^{(p)}} \log \Gamma_A(T^p_{0 \mu_1 \mu_2 \dots \mu_p})
			\frac{1}{iV^{(q)}} \log \Gamma_B(T^q_{01 \dots \nu_q})	\nonumber	\\
			&\times \mathrm{sgn}
				\begin{pmatrix}
					0 & \mu_1 & \mu_2 & \mu_3 & \dots & \mu_p & 1 & \nu_2 & \dots & \nu_q	\\
					1 & 0 & 2 & 3 & \dots & \dots & \dots & \dots & \dots & p+q
				\end{pmatrix}	\nonumber	\\
			=&-\sum_{2 \leq \mu_1 < \mu_2 < \dots < \mu_p \leq p+q}
				\frac{V^{(p+q)}}{p+q+1} \frac{1}{iV^{(p)}} \log \Gamma_A(T^p_{0 \mu_1 \mu_2 \dots \mu_p})
			\frac{1}{iV^{(q)}} \log \Gamma_B(T^q_{01 \dots \nu_q})	\nonumber	\\
			&\times \mathrm{sgn}
				\begin{pmatrix}
					0 & \mu_1 & \mu_2 & \mu_3 & \dots & \mu_p & 1 & \nu_2 & \dots & \nu_q	\\
					0 & 1 & 2 & 3 & \dots & \dots & \dots & \dots & \dots & p+q
				\end{pmatrix}	\nonumber	\\
			=&- P_4.
		\end{align}
	\item $P_6$  is a part of the summation where $\mu_0 = 0, \ \nu_1 = 1$, and $\nu_0 \neq 0$, and given by 
		\begin{align}
			P_6 =&\sum_{2 \leq \mu_1 < \mu_2 < \dots < \mu_p \leq p+q} \sum_{\nu_0 \in \{ \mu_1, \mu_2, \dots , \mu_p \}}
				\frac{V^{(p+q)}}{p+q+1} \frac{1}{iV^{(p)}} \log \Gamma_A(T^p_{0 \mu_1 \mu_2 \dots \mu_p})
			\frac{1}{iV^{(q)}} \log \Gamma_B(T^q_{\nu_0 1 \dots \nu_q})	\nonumber	\\
			&\times \mathrm{sgn}
				\begin{pmatrix}
					0 & \mu_1 & \mu_2 & \mu_3 & \dots & \mu_p & 1 & \nu_2 & \dots & \nu_q	\\
					0 & 1 & 2 & 3 & \dots & \dots & \dots & \dots & \dots & p+q
				\end{pmatrix}.
		\end{align}

	\item $P_7$  is a part of the summation where $\mu_0 = 1, \ \nu_1 = 0$, and $\nu_0 \neq 1$, and given by
		\begin{align}
			P_7=&\sum_{2 \leq \mu_1 < \mu_2 < \dots < \mu_p \leq p+q} \sum_{\nu_0 \in \{ \mu_2, \mu_3, \dots, \mu_p \}}
				\frac{V^{(p+q)}}{p+q+1} \frac{1}{iV^{(p)}} \log \Gamma_A(T^p_{1 \mu_1 \mu_2 \dots \mu_p})
			\frac{1}{iV^{(q)}} \log \Gamma_B(T^q_{\nu_0 0 \dots \nu_q})	\nonumber	\\
			&\times \mathrm{sgn}
				\begin{pmatrix}
					1 & \mu_1 & \mu_2 & \mu_3 & \dots & \mu_p & 0 & \nu_2 & \dots & \nu_q	\\
					0 & 1 & 2 & 3 & \dots & \dots & \dots & \dots & \dots & p+q
				\end{pmatrix}.	\nonumber	\\
	\mbox{Then, }			P_6|_{0 \leftrightarrow 1} =&\sum_{2 \leq \mu_1 < \mu_2 < \dots < \mu_p \leq p+q}
				\sum_{\nu_0 \in \{ \mu_1, \mu_2, \dots , \mu_p \}}
				\frac{V^{(p+q)}}{p+q+1} \frac{1}{iV^{(p)}} \log \Gamma_A(T^p_{1 \mu_1 \mu_2 \dots \mu_p})
			\frac{1}{iV^{(q)}} \log \Gamma_B(T^q_{\nu_0 0 \dots \nu_q})	\nonumber	\\
			&\times \mathrm{sgn}
				\begin{pmatrix}
					1 & \mu_1 & \mu_2 & \mu_3 & \dots & \mu_p & 0 & \nu_2 & \dots & \nu_q	\\
					1 & 0 & 2 & 3 & \dots & \dots & \dots & \dots & \dots & p+q
				\end{pmatrix}	\nonumber	\\
			=&-\sum_{2 \leq \mu_1 < \mu_2 < \dots < \mu_p \leq p+q}
				\sum_{\nu_0 \in \{ \mu_1, \mu_2, \dots , \mu_p \}}
				\frac{V^{(p+q)}}{p+q+1} \frac{1}{iV^{(p)}} \log \Gamma_A(T^p_{1 \mu_1 \mu_2 \dots \mu_p})
			\frac{1}{iV^{(q)}} \log \Gamma_B(T^q_{\nu_0 0 \dots \nu_q})	\nonumber	\\
			&\times \mathrm{sgn}
				\begin{pmatrix}
					1 & \mu_1 & \mu_2 & \mu_3 & \dots & \mu_p & 0 & \nu_2 & \dots & \nu_q	\\
					0 & 1 & 2 & 3 & \dots & \dots & \dots & \dots & \dots & p+q
				\end{pmatrix}	\nonumber	\\
			=&-P_7, \nonumber \\
	\mbox{and }		P_7|_{0\leftrightarrow 1}=& \sum_{2 \leq \mu_1 < \mu_2 < \dots < \mu_p \leq p+q}
				\sum_{\nu_0 \in \{ \mu_1, \mu_2, \dots , \mu_p \}}
				\frac{V^{(p+q)}}{p+q+1} \frac{1}{iV^{(p)}} \log \Gamma_A(T^p_{0 \mu_1 \mu_2 \dots \mu_p})
			\frac{1}{iV^{(q)}} \log \Gamma_B(T^q_{\nu_0 1 \dots \nu_q})	\nonumber	\\
			&\times \mathrm{sgn}
				\begin{pmatrix}
					0 & \mu_1 & \mu_2 & \mu_3 & \dots & \mu_p & 1 & \nu_2 & \dots & \nu_q	\\
					1 & 0 & 2 & 3 & \dots & \dots & \dots & \dots & \dots & p+q
				\end{pmatrix}	\nonumber	\\
			=&-\sum_{2 \leq \mu_1 < \mu_2 < \dots < \mu_p \leq p+q}
				\sum_{\nu_0 \in \{ \mu_1, \mu_2, \dots , \mu_p \}}
				\frac{V^{(p+q)}}{p+q+1} \frac{1}{iV^{(p)}} \log \Gamma_A(T^p_{0 \mu_1 \mu_2 \dots \mu_p})
			\frac{1}{iV^{(q)}} \log \Gamma_B(T^q_{\nu_0 1 \dots \nu_q})	\nonumber	\\
			&\times \mathrm{sgn}
				\begin{pmatrix}
					0 & \mu_1 & \mu_2 & \mu_3 & \dots & \mu_p & 1 & \nu_2 & \dots & \nu_q	\\
					0 & 1 & 2 & 3 & \dots & \dots & \dots & \dots & \dots & p+q
				\end{pmatrix}	\nonumber	\\
			=& - P_6
		\end{align}
	\item $P_8$  is a part of the summation where $\nu_1 = 0, \ \nu_2 = 1, \ \mu_0 \geq 2$, and given by
		\begin{align}
			P_8=&\sum_{2 \leq \mu_0 < \mu_1 < \dots < \mu_p \leq p+q} \sum_{\nu_0 \in \{ \mu_0 , \mu_1, \dots, \mu_p \}}
				\frac{V^{(p+q)}}{p+q+1} \frac{1}{iV^{(p)}} \log \Gamma_A(T^p_{\mu_0 \mu_1 \mu_2 \dots \mu_p})
			\frac{1}{iV^{(q)}} \log \Gamma_B(T^q_{\nu_0 0 1 \dots \nu_q})	\nonumber	\\
			&\times \mathrm{sgn}
				\begin{pmatrix}
					\mu_0 & \mu_1 & \mu_2 & \dots & \mu_p & 0 & 1 & \nu_2 & \dots & \nu_q	\\
					0 & 1 & 2 & 3 & \dots & \dots & \dots & \dots & \dots & p+q
				\end{pmatrix}.	\nonumber	\\
	\mbox{Then, }		P_8|_{0\leftrightarrow1}=&\sum_{2 \leq \mu_0 < \mu_1 < \dots < \mu_p \leq p+q} \sum_{\nu_0 \in \{ \mu_0 , \mu_1, \dots, \mu_p \}}
				\frac{V^{(p+q)}}{p+q+1} \frac{1}{iV^{(p)}} \log \Gamma_A(T^p_{\mu_0 \mu_1 \mu_2 \dots \mu_p})
			\frac{1}{iV^{(q)}} \log \Gamma_B(T^q_{\nu_0 1 0 \dots \nu_q})	\nonumber	\\
			&\times \mathrm{sgn}
				\begin{pmatrix}
					\mu_0 & \mu_1 & \mu_2 & \dots & \mu_p & 1 & 0 & \nu_2 & \dots & \nu_q	\\
					1 & 0 & 2 & 3 & \dots & \dots & \dots & \dots & \dots & p+q
				\end{pmatrix}	\nonumber	\\
			=&-\sum_{2 \leq \mu_0 < \mu_1 < \dots < \mu_p \leq p+q} \sum_{\nu_0 \in \{ \mu_0 , \mu_1, \dots, \mu_p \}}
				\frac{V^{(p+q)}}{p+q+1} \frac{1}{iV^{(p)}} \log \Gamma_A(T^p_{\mu_0 \mu_1 \mu_2 \dots \mu_p})
				\nonumber	\\
			&\times \frac{1}{iV^{(q)}} \log \Gamma_B(T^q_{\nu_0 0 1 \dots \nu_q})
				\mathrm{sgn}
				\begin{pmatrix}
					\mu_0 & \mu_1 & \mu_2 & \dots & \mu_p & 0 & 1 & \nu_2 & \dots & \nu_q	\\
					0 & 1 & 2 & 3 & \dots & \dots & \dots & \dots & \dots & p+q
				\end{pmatrix}	\nonumber	\\
			=& - P_8
		\end{align}
\end{enumerate}
Note that we have utilized the fact that link variables $\Gamma_A$ and $\Gamma_B$ satisfy
\begin{align}
	\Gamma_A(T^p) =& \Gamma^{-1}_A(-T^p),	\\
	\Gamma_B(T^q) =& \Gamma^{-1}_A(-T^q).
\end{align}
From these results, we obtain
\begin{align}
\log(\Gamma^{-1}_{A\wedge B}(T^{p+q}_{012\dots p+q}))
	=&-(P_1 + P_2 + \cdots + P_8)	\nonumber	\\
	=&(P_1 + P_2 + \cdots + P_8)_{0 \leftrightarrow 1}			\nonumber \\
	=& \log(\left. \Gamma_{A \wedge B}(T^{p+q}_{012\dots p+q}) \right|_{0 \leftrightarrow 1}).
\end{align}
Thus, we have
\begin{align}
	\Gamma^{-1}_{A\wedge B}(T^{p+q}_{012\dots p+q}) = \left. \Gamma_{A \wedge B}(T^{p+q}_{012\dots p+q}) \right|_{0 \leftrightarrow 1} .
\end{align}

We can discretize more general wedge products $A \wedge B \wedge C \wedge \cdots$ by link variables $\Gamma_{A \wedge B \wedge C \wedge \cdots}$
given by straightforward generalizations of \eqref{LinkAB}.
In the above proof, we have utilized only two facts that $\Gamma_A$ and $\Gamma_B$ satisfy $\Gamma_{A , B}(T) = \Gamma^{-1}_{A , B}(-T)$ and
that a wedge product $A \wedge B$ is discretized by a link variable \eqref{LinkAB}.
We can thus inductively show that the link variables $\Gamma_{A\wedge B\wedge C \wedge \cdots}$ satisfy
$\Gamma_{A\wedge B\wedge C\wedge \dots}(T) = \Gamma_{A\wedge B\wedge C\wedge \dots}^{-1}(-T)$ using this result.

For example, we consider the wedge product $A\wedge B \wedge C$ represented by the link variable
$\Gamma_{A\wedge B\wedge C}(T^{p+q+r})$,
where $C$ is an $r$-form.
$\Gamma_{A\wedge B\wedge C}(T^{p+q+r})$ is given by the extension of \eqref{LinkAB} as follows:
\begin{align}
	&\Gamma_{A\wedge B\wedge C}(T^{p+q+r})	\nonumber	\\
	\equiv& \exp \left[ i \sum_{T^p < T^{p+q+r}} \sum_{\substack{T^q < T^{p+q+r} \\ T^q \cap T^p = T^0}}
		\sum_{\substack{T^r< T^{p+q+r} \\ T^r \cap (T^p \cup T^q) = T'^0}} \right.	\nonumber	\\
	&\left. \times \frac{V^{(p+q+r)}}{p+q+r+1} \frac{1}{p+q+1} \frac{1}{iV^{(p)}} \log \Gamma_A(T^p)
		\frac{1}{iV^{(q)}} \log \Gamma_B(T^q) \frac{1}{iV^{(r)}} \log \Gamma_{C}(T^r) \mathrm{sgn}(\sigma) \right]
	\label{ABC}	\\
	=& \exp \left[ \sum_{0 \leq \mu_0 < \mu_1 < \dots < \mu_p \leq p+q+r} \sum_{\nu_0 \in \{ \mu_0, \mu_1, \dots, \mu_p \}}
		\sum_{\substack{0 \leq \nu_1 < \nu_2 < \dots < \nu_q \leq p+q+r \\ \nu_1, \nu_2, \dots, \nu_q \notin \{ \mu_0, \mu_1, \dots, \mu_p \}}}
		\sum_{\rho_0 \in \{ \mu_0, \mu_1, \dots, \mu_p, \nu_1, \nu_2, \dots, \nu_q \}} \frac{iV^{(p+q+r)}}{p+q+r+1} \right.	\nonumber	\\
	&\left. \times \frac{1}{p+q+1} \frac{1}{iV^{(p)}} \log \Gamma_A(T^p_{\mu_0 \mu_1 \dots \mu_p})
		\frac{1}{iV^{(q)}} \log \Gamma_B(T^q_{\nu_0 \nu_1\dots \nu_q}) \frac{1}{iV^{(r)}} \log \Gamma_{C}(T^r_{\rho_0 \rho_1\dots \rho_r})
		\mathrm{sgn}(\sigma) \right] ,	\label{ABC_vertex}
\end{align}
where $\{ \rho_1, \rho_2, \dots, \rho_r \}
		= \{0,1,2,\dots,p+q+r \} /  \{ \mu_0, \mu_1, \dots, \mu_p, \nu_1, \nu_2, \dots, \nu_q\},$
$0 \leq \rho_1 < \rho_2 < \dots < \rho_r < p+q+r$, and
\begin{align}
	&\sigma =
		\begin{pmatrix}
		\mu_0 & \mu_1 & \cdots & \mu_p & \nu_1 & \nu_2 & \cdots & \nu_q & \rho_1 & \rho_2 & \cdots & \rho_r	\\
		0 & 1 & \cdots & \cdots & \cdots & \cdots & \cdots & \cdots & \cdots & \cdots & \cdots & p+q+r
		\end{pmatrix} .
\end{align}
$\mu_0, \mu_1, \dots, \mu_p$ are vertices of $T^p$, $\nu_0, \nu_1, \dots, \nu_q$ are those of $T^q$ and $\nu_0$ is also shared by $T^p$,
and $\rho_0, \rho_1, \dots, \rho_r$ are those of $T^r$ and $\rho_0$ is also shared by the union $T^p \cup T^q$.

In \eqref{ABC}, we choose one of the sub $p$-simplex $T^p$ of $T^{p+q+r}$, and
then choose the one of the sub $q$-simplex $T^q$ of $T^{p+q+r}$ that shares one of its vertices $T^0$ with $T^p$.
This is equivalent to choosing the one of the sub $(p+q)$-simplex $T^{p+q}$ of $T^{p+q+r}$,
and then $p$-subsimplex $T^p$ and $q$-subsimplex $T^q$ of $T^{p+q}$ so that they share one of their vertices $T^0$.
Using this fact, we rewrite the summations over subsimplex in \eqref{ABC} as
\begin{align}
	\sum_{T^p < T^{p+q+r}} \sum_{\substack{T^q < T^{p+q+r} \\ T^q \cap T^p = T^0}}
		\sum_{\substack{T^r< T^{p+q+r} \\ T^r \cap (T^p \cup T^q) = T'^0}}
	=& \sum_{T^{p+q} < T^{p+q+r}} \sum_{T^p < T^{p+q}} \sum_{\substack{T^q < T^{p+q} \\ T^q \cap T^p = T^0}}
		\sum_{\substack{T^r< T^{p+q+r} \\ T^r \cap T^{p+q} = T'^0}}.	\label{rewrite_sum}
\end{align}
Using \eqref{rewrite_sum}, we obtain
\begin{align}
	&\Gamma_{A\wedge B\wedge C}(T^{p+q+r})	\nonumber	\\
	=& \exp \left[ \sum_{T^{p+q} < T^{p+q+r}} \sum_{T^p < T^{p+q}} \sum_{\substack{T^q < T^{p+q} \\ T^q \cap T^p = T^0}}
		\sum_{\substack{T^r< T^{p+q+r} \\ T^r \cap T^{p+q} = T'^0}} \right.	\nonumber	\\
	&\left. \times \frac{iV^{(p+q+r)}}{p+q+r+1} \frac{1}{p+q+1} \frac{1}{iV^{(p)}} \log \Gamma_A(T^p)
		\frac{1}{iV^{(q)}} \log \Gamma_B(T^q) \frac{1}{iV^{(r)}} \log \Gamma_{C}(T^r) \mathrm{sgn}(\sigma) \right]
		\nonumber	\\
	=& \exp \left[ \frac{iV^{(p+q+r)}}{p+q+r+1}
		\sum_{T^{p+q} < T^{p+q+r}} \frac{\mathrm{sgn}(\sigma'(T^p))}{iV^{(p+q)}} \sum_{T^p < T^{p+q}}
		\sum_{\substack{T^q < T^{p+q} \\ T^q \cap T^p = T^0}}
		\frac{iV^{(p+q)}}{p+q+1} \frac{1}{iV^{(p)}} \log \Gamma_A(T^p) \right.	\nonumber	\\
	&\left. \times \frac{1}{iV^{(q)}} \log \Gamma_B(T^q) \sum_{\substack{T^r< T^{p+q+r} \\ T^r \cap T^{p+q} = T'^0}}
		\frac{1}{iV^{(r)}} \log \Gamma_{C}(T^r) \mathrm{sgn}(\sigma'') \right]
		\nonumber	\\
	=& \exp \left[ \frac{iV^{(p+q+r)}}{p+q+r+1}
		\sum_{T^{p+q} < T^{p+q+r}} \frac{1}{iV^{(p+q)}} \log \Gamma_{A \wedge B}(T^{p+q}) \sum_{\substack{T^r< T^{p+q+r} \\ T^r \cap T^{p+q} = T'^0}}
		\frac{1}{iV^{(r)}} \log \Gamma_{C}(T^r) \mathrm{sgn}(\sigma'') \right]	\nonumber\\
	\equiv& \Gamma_{(A\wedge B)\wedge C}(T^{p+q+r}).	\label{ABC_result}
\end{align}
Since $\Gamma_{A\wedge B}(T^{p+q})$ satisfies $\Gamma_{A\wedge B}(T^{p+q}) = \Gamma^{-1}_{A \wedge B}(-T^{p+q})$
as we have already shown, we have
\begin{align}
	\Gamma_{(A \wedge B) \wedge C}(T^{p+q+r}) = \Gamma^{-1}_{(A \wedge B) \wedge C}(-T^{p+q+r}).
\end{align}
Thus, we have proven $\Gamma_{A\wedge B\wedge C}(T) = \Gamma_{A\wedge B\wedge C}^{-1}(-T)$.
In the second equality of \eqref{ABC_result}, we have used
\begin{align}
	\mathrm{sgn}(\sigma) = \mathrm{sgn}(\sigma') \mathrm{sgn}(\sigma''),
\end{align}
and this can be easily shown as follows.
$\sigma, \sigma'$, and $\sigma''$ can be written in terms of $(\mu_0, \mu_1, \dots , \mu_{p-1}, \mu_p)$,
$(\nu_1, \nu_2, \dots, \nu_{q-1}, \nu_q)$, and $(\rho_1, \rho_2, \dots, \rho_{r-1}, \rho_r)$,
the vertices of $T^p, T^q$, and $T^r$, respectively, as
\begin{align}
	&\sigma =
		\begin{pmatrix}
		\mu_0 & \mu_1 & \cdots & \mu_p & \nu_1 & \nu_2 & \cdots & \nu_q & \rho_1 & \rho_2 & \cdots & \rho_r	\\
		0 & 1 & \cdots & \cdots & \cdots & \cdots & \cdots & \cdots & \cdots & \cdots & \cdots & p+q+r
		\end{pmatrix},	\\
	&\sigma' = \begin{pmatrix}
		\mu_0 & \mu_1 & \cdots & \mu_p & \nu_1 & \nu_2 & \cdots & \nu_q	\\
		0 & 1 & \cdots &\hat{\rho}_1  & \cdots & \hat{\rho}_r & \cdots &p+q+r
		\end{pmatrix},
\end{align}
and
\begin{align}
	&\sigma'' = \begin{pmatrix}
		0 & 1 & \cdots &\hat{\rho}_1  & \cdots & \hat{\rho}_r & \cdots &p+q+r & \rho_1 & \rho_2 & \cdots & \rho_r	\\
		0 & 1 & \cdots & \cdots & \cdots & \cdots & \cdots & \cdots & \cdots & \cdots & \cdots & p+q+r
		\end{pmatrix}.
\end{align}
Using the fact
\begin{align}
	&\{ \rho_1, \rho_2, \dots, \rho_r \}
		= \{0,1,2,\dots,p+q+r \}/ \{ \mu_0, \mu_1, \dots, \mu_p, \nu_1, \nu_2, \dots, \nu_q\},
\end{align}
we can rewrite $\sigma$ as a product of permutation of $(\mu_0, \mu_1,  \dots, \mu_p, \nu_1, \nu_2, \dots \nu_{q-1}, \nu_q)$, and that of $(\rho_1, \rho_2, \dots \rho_{r-1}, \rho_r)$:
\begin{align}
	\sigma
		=& \sigma'\times \sigma''.
\end{align}

\section{$\Gamma^{-1}_{dA}(T^{p+1})=\Gamma_{dA}(-T^{p+1})$}
In this section, we verify that the plaquette variable $\Gamma_{dA}(T^{p+1})$ of the exterior derivative $dA$ of a $p$-form $A$ satisfies
\begin{align}
	\Gamma^{-1}_{dA}(T^{p+1}) = \Gamma_{dA}(-T^{p+1}). \label{diff}
\end{align}
Since $T^{p}_{012 \dots p+1} = - T^{p}_{102 \dots p+1}$, it is enough to show
\begin{align}
	\Gamma^{-1}_{dA}(T^{p+1}_{012\dots p+1}) = \Gamma_{dA}(T^{p+1}_{012\dots p+1})|_{0 \leftrightarrow 1}.
	\label{GammaAInverse}
\end{align}

First, $\Gamma_{dA}(T^{p+1})$ is given by
\begin{align}
	&\Gamma_{dA}(T^{p+1})	\nonumber	\\
	=& \exp \left[ i \sqrt{\frac{p+2}{2(p+1)}} \sum_{T^p \subset \partial T^{p+1}} \frac{1}{i} \log \Gamma_{A}(T^p) \right]
		\nonumber	\\
	=& \exp \left[ i \sqrt{\frac{p+2}{2(p+1)}} \sum_{k =0}^{p+1} \frac{(-1)^k}{i} \log \Gamma_{A}(T^p_{012\dots \hat{k} \dots p+1}) \right]
		\nonumber	\\
	=& \exp \left[ i \sqrt{\frac{p+2}{2(p+1)}} \left\{  \sum_{k =2}^{p+1} \frac{(-1)^k}{i} \log \Gamma_{A}(T^p_{012\dots \hat{k} \dots p+1})
		+ \frac{1}{i} \log \Gamma_A(T^p_{123 \dots p+1}) - \frac{1}{i} \log \Gamma_A(T^p_{023 \dots p+1}) \right\} \right],
\end{align}
where $\Gamma_A(T^p)$ is a link variable of $A$ and satisfies $\Gamma_A(-T^p) = \Gamma^{-1}_A(T^p)$.
Next, exchanging the label of vertices as $0 \leftrightarrow 1$, we obtain
\begin{align}
&\Gamma_{dA}(T^{p+1}_{012\dots p+1})|_{0 \leftrightarrow 1} \nonumber \\
	=& \exp \left[ i \sqrt{\frac{p+2}{2(p+1)}} \left\{  \sum_{k =2}^{p+1} \frac{(-1)^k}{i} \log \Gamma_{A}(T^p_{012\dots \hat{k} \dots p+1})
		+ \frac{1}{i} \log \Gamma_A(T^p_{123 \dots p+1}) - \frac{1}{i} \log \Gamma_A(T^p_{023 \dots p+1}) \right\} \right]_{0 \leftrightarrow 1}
		\nonumber	\\
	=& \exp \left[ i \sqrt{\frac{p+2}{2(p+1)}} \left\{  \sum_{k =2}^{p+1} \frac{(-1)^k}{i} \log \Gamma_{A}(T^p_{102\dots \hat{k} \dots p+1})
		+ \frac{1}{i} \log \Gamma_A(T^p_{023 \dots p+1}) - \frac{1}{i} \log \Gamma_A(T^p_{123 \dots p+1}) \right\} \right]
		\nonumber	\\
	=& \exp \left[ i \sqrt{\frac{p+2}{2(p+1)}} \left\{  - \frac{1}{i} \log \Gamma_A(T^p_{123 \dots p+1}) 
		+ \frac{1}{i} \log \Gamma_A(T^p_{023 \dots p+1})
		- \sum_{k =2}^{p+1} \frac{(-1)^k}{i} \log \Gamma_{A}(T^p_{012\dots \hat{k} \dots p+1}) \right\} \right]
		\nonumber	\\
	=& \exp \left[ - i \sqrt{\frac{p+2}{2(p+1)}} \sum_{k = 0}^{p+1} \frac{(-1)^k}{i} \log \Gamma_{A}(T^p_{012\dots \hat{k} \dots p+1}) \right]
		\nonumber	\\
	=& \Gamma^{-1}_{dA}(T^p_{012 \dots p+1}).
\end{align}
From this result, we can see that $\Gamma_{dA}(T^p)$ satisfies \eqref{GammaAInverse}.

\section{Leibniz rule on simplex}\label{Leibniz_rule}
It has been pointed out that the Leibniz rule is violated on the lattice.
For example, we consider $\partial(f(x) g(x))$, where $f(x)$ and $g(x)$ are scalar functions.
In the continuum, we have
\begin{align}
	\partial(f(x)g(x)) = \partial f(x) g(x) + f(x) \partial g(x).	\label{LeibnizRuleinContinuum}
\end{align}
On the lattice, derivatives are replaced with differences as
\begin{align}
	\partial f(x) = \lim_{dx\to 0} \frac{f(x+dx) - f(x)}{dx} \to \frac{f(x+a \hat{\mu}) - f(x)}{a},
\end{align}
where $a$ is a lattice spacing and $\hat{\mu}$ is a unit vector in $x$ direction.
Discretizing the both sides of \eqref{LeibnizRuleinContinuum} by using this replacement, we obtain
\begin{align}
	\partial (f(x) g(x))
	\to& \frac{f(x+a \hat{\mu}) g(x+a \hat{\mu}) - f(x) g(x)}{a}	\nonumber	\\
	=& \frac{f(x+a \hat{\mu}) - f(x)}{a}g(x) + \frac{g(x+a \hat{\mu}) - g(x)}{a}f(x)	\nonumber	\\
	&+a \frac{f(x+a \hat{\mu}) - f(x)}{a} \frac{g(x+a \hat{\mu}) - g(x)}{a},	\label{LeibnizRuleLHS}
\end{align}
from the left-hand side and
\begin{align}
	\partial f(x) g(x) + f(x) \partial g(x) \to& \frac{1}{a} \left\{ f(x+a\hat{\mu})g(x) - f(x)g(x) + f(x)g(x+a\hat{\mu}) - f(x)g(x) \right\}	\nonumber	\\
	=& \frac{f(x+a \hat{\mu}) - f(x)}{a}g(x) + \frac{g(x+a \hat{\mu}) - g(x)}{a}f(x),	\label{LeibnizRuleRHS}
\end{align}
from the right-hand side.
\eqref{LeibnizRuleLHS} does not agree with \eqref{LeibnizRuleRHS} due to the lattice artifact that is formally of order $O(a)$.
We can thus see that we cannot naively discretize both sides of \eqref{LeibnizRuleinContinuum}.

However, in the special case where $\partial f = 0$ or $\partial g=0$ holds, we can ignore the lattice artifact and discretize both sides of \eqref{LeibnizRuleinContinuum}.

This fact can be generalized to the case of tensors.
For instance, we consider the discretization of
\begin{align}
d (df \cdot g) = - df \wedge dg	\label{dfdg}
\end{align}
on the 2-simplex $T^2_{012}$.
Supposing that the length of all edges of $T^2_{012}$ is $a$, the discretization gives
\begin{align}
	(\mathrm{LHS \ of \ \eqref{dfdg}}) 
	\to&\sum_{T^1 \subset \partial T^2_{012}} \sqrt{\frac{3}{4}}
		\left( \sum_{T^0 \subset \partial T^1} f(T^0) \right) \left( \frac{1}{2} \sum_{T^0<T^1} g(T^0) \right),	\label{dfdg_LHS}	\\
	(\mathrm{RHS \ of \ \eqref{dfdg}})
	\to& - \frac{\sqrt{3}a^2}{4} \sum_{0 \leq \mu_0 < \mu_1 \leq 2} \sum_{\nu_0 \in \{ \mu_0, \mu_1\}} \frac{1}{3}
		\frac{f(T^0_{\mu_1}) - f(T^0_{\mu_0})}{a} \frac{g(T^0_{\nu_1}) - g(T^0_{\nu_0})}{a} \mathrm{sgn}(\sigma),
		\label{dfdg_RHS}
\end{align}
where $\nu_1 \in \{0,1,2 \}/\{ \mu_0, \mu_1 \}$, and
\begin{align}
	\sigma =&	\begin{pmatrix}
			\mu_0 & \mu_1 & \nu_1	\\
			0 & 1 & 2
			\end{pmatrix}.
\end{align}
After some calculation, we  obtain from \eqref{dfdg_LHS}, 
\begin{align}
	&\sum_{T^1 \subset \partial T^2_{012}} \sqrt{\frac{3}{4}} \left( \sum_{T^0 \subset \partial T^1} f(T^0) \right) \left( \frac{1}{2} \sum_{T^0<T^1} g(T^0) \right)
		\nonumber	\\
	=& \sqrt{\frac{3}{4}} \frac{1}{2} \left[ (f(T^0_2) - f(T^0_1)) \left( g(T^0_1) + g(T^0_2) \right)
		- ({f(T^0_2) - f(T^0_0)}) \left( g(T^0_2) + g(T^0_0) \right) \right. \nonumber	\\
	+&\left. ({f(T^0_1) - f(T^0_0)}) \left( g(T^0_1) + g(T^0_0) \right) \right]	\nonumber	\\
	=& \sqrt{\frac{3}{4}} \frac{1}{2} \left[ {(f(T^0_2) - f(T^0_1))} \left( g(T^0_1) + g(T^0_2) \right)
		- ({f(T^0_2) - f(T^0_0)}) \left( g(T^0_2) + g(T^0_0) \right) \right. \nonumber	\\
	+&\left. ({f(T^0_1) - f(T^0_0)}) \left( g(T^0_1) + g(T^0_0) \right) \right]	\nonumber	\\
	=& \frac{\sqrt{3}}{4} \left[ f(T^0_0)(g(T^0_2) - g(T^0_1)) + f(T^0_1)(g(T^0_0) - g(T^0_2)) + f(T^0_2)(g(T^0_1) - g(T^0_0)) \right],
		\label{dfdg_LHS_result}
\end{align}
and from \eqref{dfdg_RHS},
\begin{align}
	&- \frac{\sqrt{3}a^2}{4} \sum_{0 \leq \mu_0 < \mu_1 \leq 2} \sum_{\nu_0 \in \{ \mu_0, \mu_1\}} \frac{1}{3}
		\frac{f(T^0_{\mu_1}) - f(T^0_{\mu_0})}{a} \frac{g(T^0_{\nu_1}) - g(T^0_{\nu_0})}{a} \mathrm{sgn}(\sigma)	\nonumber	\\
	=&- \frac{\sqrt{3}}{12} \left[ (f(T^0_1) - f(T^0_0)) \left\{ (g(T^0_2) - g(T^0_1)) + (g(T^0_2) - g(T^0_0)) \right\} \right.	\nonumber	\\
	&- (f(T^0_2) - f(T^0_0)) \left\{ (g(T^0_1) - g(T^0_2)) + (g(T^0_1) - g(T^0_0)) \right\}	\nonumber	\\
	&\left. +(f(T^0_2) - f(T^0_1)) \left\{ (g(T^0_0) - g(T^0_2)) + (g(T^0_0) - g(T^0_1)) \right\} \right]	\nonumber	\\
	=&\frac{\sqrt{3}}{4} \left[ f(T^0_0)(g(T^0_2) - g(T^0_1)) + f(T^0_1)(g(T^0_0) - g(T^0_2)) + f(T^0_2)(g(T^0_1) - g(T^0_0)) \right].
		\label{dfdg_RHS_result}
\end{align}
As \eqref{dfdg_LHS_result} agrees with \eqref{dfdg_RHS_result}, we see that \eqref{dfdg} can be discretized on the simplex
and that the Leibniz rule is not violated on the lattice.

For another example, we will discretize the both sides of
\begin{align}
	d(d\Lambda_3\wedge B_{2}) = - d\Lambda_3 \wedge dB_{2}	\label{dLambdaWedgeB}
\end{align}
on a 7-simplex $T^7_{012 \dots 7}$ and show that results are the same.
First, the left-hand side of \eqref{dLambdaWedgeB} is discretized as
\begin{align}
	&d(d \Lambda_3 \wedge B_2)	\nonumber	\\
	\to&\frac{\sqrt{8}}{\sqrt{14}} \sum_{T^6 \subset \partial T^{7}} \sum_{T^4<T^6} \sum_{T^3 \subset \partial T^4}
		\sum_{\substack{T^{2} < T^6 \\ T^{2} \cap T^{4} = T^0}} \frac{V^{(6)}}{7} \frac{1}{iV^{(4)}} \sqrt{\frac{5}{8}} \log \Gamma_\Lambda(T^3)
		\frac{1}{iV^{(2)}} \log \Gamma_{B}(T^{2}) \mathrm{sgn} (\sigma)	\nonumber	\\
	=&\frac{\sqrt{8}}{\sqrt{14}} \sum_{r=0}^7 \sum_{0 \leq \nu_1 < \nu_2 \leq 7}
		\left[ \sum_{0 \leq q < r < \nu_1 < \nu_2} + \sum_{0 \leq q < \nu_1 < r < \nu_2} + \sum_{0 \leq q < \nu_1 < \nu_2 < r} \right.
		\nonumber	\\
	&- \sum_{r < q < \nu_1 < \nu_2} - \sum_{\nu_1 < q < r < \nu_2} - \sum_{\nu_1 < q < \nu_2 < r}
		+ \sum_{r < \nu_1 < q < \nu_2} + \sum_{\nu_1 < r < q  < \nu_2} + \sum_{\nu_1 < \nu_2 < q < r}
		\nonumber	\\
	&\left. - \sum_{\leq r < \nu_1 < \nu_2 < q \leq 7 } - \sum_{\nu_1< r < \nu_2 < q \leq 7} - \sum_{\nu_1< \nu_2 < r < q \leq 7} \right]
		\sum_{\nu_0 \in \{ 0,1,2,\dots 7 \} / \{ r, \nu_1, \nu_2 \}} (-1)^{r+q}
		\nonumber	\\
	&\times \frac{V^{(6)}}{7} \frac{1}{iV^{(4)}} \sqrt{\frac{5}{8}} \log \Gamma_\Lambda(T^{3, \hat{r} \hat{q} \hat{\nu}_1 \hat{\nu}_2}_{012\dots 7})
		\frac{1}{iV^{(2)}} \log \Gamma_{B}(T^{2}_{\nu_0 \nu_1 \nu_2}) \mathrm{sgn} (\sigma),
		\label{D13}
\end{align}
where
\begin{align}
	\sigma =& \begin{pmatrix}
	0 & 1 & 2 & \cdots & \hat{r}, \hat{\nu}_1, \hat{\nu}_2 & \cdots & 7 & \nu_1 & \nu_2	\\
	0 & 1 & 2 & \cdots & \hat{r} & \cdots & \cdots & \cdots & 7
	\end{pmatrix},
\end{align}
$r$ and $q$ are vertex that we omit when we obtain the 6-simplex $T^6$ from $T^7_{012 \dots 7}$ and 3-simplex $T^3$
from 4-simplex $T^4$, respectively, and $T^{3, \hat{r} \hat{q} \hat{\nu}_1 \hat{\nu}_2}_{012\dots 7}$ is a 3-simplex
obtained by omitting vertices $T^0_r, T^0_q, T^0_{\nu_1}$, and $ T^0_{\nu_2}$ from $T^7_{012 \dots 7}$.

Using the fact that
\begin{align}
	\mathrm{sgn}(\sigma) = \begin{cases}
	(-1)^{\nu_1 + \nu_2-1} \ &(r<\nu_1)	\\
	(-1)^{\nu_1 + \nu_2}\ \ &(\nu_1 < r < \nu_2)	\\
	(-1)^{\nu_1 + \nu_2-1}\ \ &(\mu_1 < \mu_2 < r)
	\end{cases},
\end{align}
The right hand side of (\ref{D13}) becomes
\begin{align}
	=&\frac{\sqrt{8}}{\sqrt{14}} \sum_{r=0}^7 \sum_{0 \leq \nu_1 < \nu_2 \leq 7}
		\left[ \sum_{0 \leq q < r < \nu_1 < \nu_2} + \sum_{0 \leq q < \nu_1 < r < \nu_2} + \sum_{0 \leq q < \nu_1 < \nu_2 < r} \right.
		\nonumber	\\
	&- \sum_{r < q < \nu_1 < \nu_2} - \sum_{\nu_1 < q < r < \nu_2} - \sum_{\nu_1 < q < \nu_2 < r}
		+ \sum_{r < \nu_1 < q < \nu_2} + \sum_{\nu_1 < r < q  < \nu_2} + \sum_{\nu_1 < \nu_2 < q < r}
		\nonumber	\\
	&\left. - \sum_{\leq r < \nu_1 < \nu_2 < q \leq 7 } - \sum_{\nu_1< r < \nu_2 < q \leq 7} - \sum_{\nu_1< \nu_2 < r < q \leq 7} \right]
		\sum_{\nu_0 \in \{ 0,1,2,\dots 7 \} / \{ r, \nu_1, \nu_2 \}} (-1)^{r+q+\nu_1+\nu_2-1}
		\nonumber	\\
	&\times \frac{V^{(6)}}{7} \frac{1}{iV^{(4)}} \sqrt{\frac{5}{8}} \log \Gamma_\Lambda(T^{3, \hat{r} \hat{q} \hat{\nu}_1 \hat{\nu}_2}_{012\dots 7})
		\frac{1}{iV^{(2)}} \log \Gamma_{B}(T^{2}_{\nu_0 \nu_1 \nu_2})	\nonumber	\\
	=&\frac{\sqrt{8}}{\sqrt{14}} \sum_{r=0}^7 \sum_{0 \leq \nu_1 < \nu_2 \leq 7}
		\left[ \sum_{0 \leq q < r < \nu_1 < \nu_2} - \sum_{0 \leq q < \nu_1 < r < \nu_2} + \sum_{0 \leq q < \nu_1 < \nu_2 < r} \right.
		\nonumber	\\
	&- \sum_{r < q < \nu_1 < \nu_2} + \sum_{\nu_1 < q < r < \nu_2} - \sum_{\nu_1 < q < \nu_2 < r}
		+ \sum_{r < \nu_1 < q < \nu_2} - \sum_{\nu_1 < r < q  < \nu_2} + \sum_{\nu_1 < \nu_2 < q < r}
		\nonumber	\\
	&\left. - \sum_{\leq r < \nu_1 < \nu_2 < q \leq 7 } + \sum_{\nu_1< r < \nu_2 < q \leq 7} - \sum_{\nu_1< \nu_2 < r < q \leq 7} \right]
		(-1)^{r+q+\nu_1+\nu_2-1}
		\nonumber	\\
	&\times \frac{V^{(6)}}{7} \frac{1}{iV^{(4)}} \sqrt{\frac{5}{8}} \log \Gamma_\Lambda(T^{3, \hat{r} \hat{q} \hat{\nu}_1 \hat{\nu}_2}_{012\dots 7})
		\frac{1}{iV^{(2)}}
		\left\{ \left( \sum_{\nu_0 \in \{ 0,1,2,\dots 7 \}} \log \Gamma_{B}(T^{2}_{\nu_0 \nu_1 \nu_2}) \right)
			- \log \Gamma_{B}(T^{2}_{r \nu_1 \nu_2}) \right\}	\nonumber	\\
	=&\frac{\sqrt{8}}{\sqrt{14}} \sum_{0 \leq q < r < \nu_1 < \nu_2 \leq 7} (-1)^{r+q+\nu_1+\nu_2-1}
		\frac{V^{(6)}}{7} \frac{1}{iV^{(4)}} \sqrt{\frac{5}{8}} \log \Gamma_\Lambda(T^{3, \hat{r} \hat{q} \hat{\nu}_1 \hat{\nu}_2}_{012\dots 7})
		\frac{1}{iV^{(2)}}
		\nonumber	\\
	&\times \left[ \left\{ \left( \sum_{\nu_0 \in \{ 0,1,2,\dots 7 \}} \log \Gamma_{C}(T^{2}_{\nu_0 \nu_1 \nu_2}) \right)
			- \log \Gamma_{C}(T^{2}_{r \nu_1 \nu_2}) \right\}
			- \left\{ \left( \sum_{\nu_0 \in \{ 0,1,2,\dots 7 \}} \log \Gamma_{C}(T^{2}_{\nu_0 r \nu_2}) \right)
			- \log \Gamma_{C}(T^{2}_{\nu_1 r \nu_2}) \right\} \right.	\nonumber	\\
	&+ \left\{ \left( \sum_{\nu_0 \in \{ 0,1,2,\dots 7 \}} \log \Gamma_{C}(T^{2}_{\nu_0 r \nu_1}) \right)
			- \log \Gamma_{C}(T^{2}_{\nu_2 r \nu_1}) \right\}
		- \left\{ \left( \sum_{\nu_0 \in \{ 0,1,2,\dots 7 \}} \log \Gamma_{C}(T^{2}_{\nu_0 \nu_1 \nu_2}) \right)
			- \log \Gamma_{C}(T^{2}_{q \nu_1 \nu_2}) \right\}	\nonumber	\\
	&+ \left\{ \left( \sum_{\nu_0 \in \{ 0,1,2,\dots 7 \}} \log \Gamma_{C}(T^{2}_{\nu_0 q \nu_2}) \right)
			- \log \Gamma_{C}(T^{2}_{\nu_1 q \nu_2}) \right\}
		-\left\{ \left( \sum_{\nu_0 \in \{ 0,1,2,\dots 7 \}} \log \Gamma_{C}(T^{2}_{\nu_0 q \nu_1}) \right)
			- \log \Gamma_{C}(T^{2}_{\nu_2 q \nu_1}) \right\}	\nonumber	\\
	&+ \left\{ \left( \sum_{\nu_0 \in \{ 0,1,2,\dots 7 \}} \log \Gamma_{C}(T^{2}_{\nu_0 r \nu_2}) \right)
			- \log \Gamma_{C}(T^{2}_{q r \nu_2}) \right\}
		- \left\{ \left( \sum_{\nu_0 \in \{ 0,1,2,\dots 7 \}} \log \Gamma_{C}(T^{2}_{\nu_0 q \nu_2}) \right)
			- \log \Gamma_{C}(T^{2}_{r q \nu_2}) \right\}	\nonumber	\\
	&+ \left\{ \left( \sum_{\nu_0 \in \{ 0,1,2,\dots 7 \}} \log \Gamma_{C}(T^{2}_{\nu_0 q r}) \right)
			- \log \Gamma_{C}(T^{2}_{\nu_2 q r}) \right\}
		-\left\{ \left( \sum_{\nu_0 \in \{ 0,1,2,\dots 7 \}} \log \Gamma_{C}(T^{2}_{\nu_0 r \nu_1}) \right)
			- \log \Gamma_{C}(T^{2}_{q r \nu_1}) \right\}	\nonumber	\\
	&+\left\{ \left( \sum_{\nu_0 \in \{ 0,1,2,\dots 7 \}} \log \Gamma_{C}(T^{2}_{\nu_0 q \nu_1}) \right)
			- \log \Gamma_{C}(T^{2}_{r q \nu_1}) \right\}
		-\left\{ \left( \sum_{\nu_0 \in \{ 0,1,2,\dots 7 \}} \log \Gamma_{C}(T^{2}_{\nu_0 q r}) \right)
			- \log \Gamma_{C}(T^{2}_{\nu_1 q r}) \right\}	\nonumber	\\
	=&\sqrt{\frac{{8}}{{14}}} \sum_{0 \leq q < r < \nu_1 < \nu_2 \leq 7} (-1)^{r+q+\nu_1+\nu_2}
		\frac{V^{(6)}}{7} \frac{1}{iV^{(4)}} \sqrt{\frac{5}{8}} \log \Gamma_\Lambda(T^{3, \hat{r} \hat{q} \hat{\nu}_1 \hat{\nu}_2}_{012\dots 7})
		\frac{3}{iV^{(2)}}
		\nonumber	\\
	&\times \left[ \log \Gamma_B(T^2_{r \nu_1 \nu_2}) - \log \Gamma_{B}(T^{2}_{q \nu_1 \nu_2})
		+ \log \Gamma_B(T^2_{r q \nu_1}) - \log \Gamma_B(T^2_{r q \nu_2}) \right].	\label{dLambdaWedgedB_LHS}
\end{align}

Next, the right-hand side of \eqref{dLambdaWedgeB} is discretized as
\begin{align}
	&d\Lambda \wedge dB	\nonumber	\\
	\to& - \frac{V^{(7)}}{8} \sum_{T^4 < T^{7}} \sum_{T^3 \subset \partial T^4}
		\sum_{\substack{T'^{3} < T^6 \\ T'^{3} \cap T^{4} = T^0}} \sum_{T'^{2} \subset \partial T'^{3}}
		\frac{1}{iV^{(4)}} \sqrt{\frac{5}{8}} \log \Gamma_\Lambda(T^2)
		\frac{1}{iV'^{(3)}} \sqrt{\frac{4}{6}}
		\log \Gamma_{B}(T'^{2}) \mathrm{sgn} (\sigma')	\nonumber	\\
	=&\frac{V^{(7)}}{8} \sum_{0 \leq < \nu_1 < \nu_2 < \nu_3 \leq 7} \left[ \sum_{0 \leq r < \nu_1} - \sum_{\nu_1 < r < \nu_2}
		+ \sum_{\nu_2 < r < \nu_3} - \sum_{\nu_3 < r \leq 7} \right] (-1)^r
		\frac{1}{iV^{(4)}} \sqrt{\frac{5}{8}} \log \Gamma_{\Lambda}(T^{3, \hat{r} \hat{\nu}_1 \hat{\nu}_2 \hat{\nu}_3}_{012\dots 7})	\nonumber	\\
	&\times \sum_{\nu_{0} \in \{ 0,1,2, \dots , 7 \} / \{\nu_1, \nu_2, \nu_3\}} \sum_{q=0}^{3} \frac{(-1)^q}{iV^{(3)}} \sqrt{\frac{{4}}{{6}}}
		\log \Gamma_B (T^{2, \hat{\nu}_q}_{\nu_0 \nu_1 \nu_2 \nu_{3}}) \mathrm{sgn}(\sigma'),
		\label{D17}
\end{align}
where
\begin{align}
	\sigma' = \begin{pmatrix}
	0 & 1 & 2 & \cdots & \hat{\nu}_1, \hat{\nu}_2, \hat{\nu}_3 & \cdots & 7 & \nu_1 & \nu_2 & \nu_3	\\
	0 & 1 & 2 & \cdots & \cdots & \cdots & \cdots & \cdots & \cdots & 7
	\end{pmatrix},
\end{align}
$\nu_1$, $\nu_2$ and $\nu_3$ are the vertices of $T'^3$ that are not shared by $T^4$,
and $r$ and $\nu_q$ are the vertex which we omit when we obtain $T^3$ from $T^4$ and $T'^2$ from $T'^3$, respectively.
Using the fact that
\begin{align}
	\mathrm{sgn}(\sigma') = (-1)^{\nu_1 + \nu_2 + \nu_3},
\end{align}
the right hand side of (\ref{D17}) becomes
\begin{align}
	&\frac{V^{(7)}}{8} \sum_{0 \leq < \nu_1 < \nu_2 < \nu_3 \leq 7} \left[ \sum_{0 \leq r < \nu_1} - \sum_{\nu_1 < r < \nu_2}
		+ \sum_{\nu_2 < r < \nu_3} - \sum_{\nu_3 < r \leq 7} \right] (-1)^r
		\frac{1}{iV^{(4)}} \sqrt{\frac{5}{8}} \log \Gamma_{\Lambda}(T^{3, \hat{r} \hat{\nu}_1 \hat{\nu}_2 \hat{\nu}_3}_{012\dots 7})	\nonumber	\\
	&\times \sum_{\nu_{0} \in \{ 0,1,2, \dots , 7 \} / \{\nu_1, \nu_2, \nu_3\}} \sum_{q=0}^{3} \frac{(-1)^q}{iV^{(3)}} \sqrt{\frac{{4}}{{6}}}
		\log \Gamma_B (T^{2, \hat{\nu}_q}_{\nu_0 \nu_1 \nu_2 \nu_{3}}) \mathrm{sgn}(\sigma')	\nonumber	\\
	=& \frac{V^{(7)}}{8} \sum_{0 \leq \nu_1 < \nu_2 < \nu_3 \leq 7} \left[ \sum_{0 \leq r < \nu_1} - \sum_{\nu_1 < r < \nu_2}
		+ \sum_{\nu_2 < r < \nu_3} - \sum_{\nu_3 < r \leq 7} \right] (-1)^{r+\nu_1 + \nu_2 + \nu_3}
		\frac{1}{iV^{(4)}} \sqrt{\frac{5}{8}} \log \Gamma_{\Lambda}(T^{3, \hat{r} \hat{\nu}_1 \hat{\nu}_2 \hat{\nu}_3}_{012\dots 7})	\nonumber	\\
	&\times \frac{1}{iV^{(3)}} \sqrt{\frac{{4}}{{6}}} \left[\sum_{\nu_{0} \in \{ 0,1,2, \dots , 7 \}}
		\left\{\log \Gamma_B (T^{2}_{\nu_1 \nu_2 \nu_{3}}) - \log \Gamma_B (T^{2}_{\nu_0 \nu_2 \nu_{3}})
		+ \log \Gamma_B (T^{2}_{\nu_0 \nu_1 \nu_{3}}) - \log \Gamma_B (T^{2}_{\nu_0 \nu_1 \nu_2}) \right\} \right.	\nonumber	\\
	&\biggl. - 3 \log \Gamma_B(T^2_{\nu_1 \nu_2 \nu_3}) + \log \Gamma_B(T^2_{\nu_1 \nu_2 \nu_3}) - \log \Gamma_B(T^2_{\nu_2 \nu_1 \nu_3})
		+ \log \Gamma_B(T^2_{\nu_3 \nu_1 \nu_2}) \biggr]	\nonumber	\\
	=& \frac{V^{(7)}}{8} \left[ \sum_{0 \leq r < \nu_1 < \nu_2 < \nu_3 \leq 7} - \sum_{0 \leq \nu_1 < r < \nu_2 < \nu_3 \leq 7}
		+ \sum_{0 \leq \nu_1 < \nu_2 < r < \nu_3 \leq 7} - \sum_{0 \leq \nu_1 < \nu_2 < \nu_3 < r \leq 7} \right] (-1)^{r+\nu_1 + \nu_2 + \nu_3}
		\nonumber	\\
	&\times \frac{1}{iV^{(4)}} \sqrt{\frac{5}{8}} \log \Gamma_{\Lambda}(T^{3, \hat{r} \hat{\nu}_1 \hat{\nu}_2 \hat{\nu}_3}_{012\dots 7})
		\frac{1}{iV^{(3)}} \sqrt{\frac{{4}}{{6}}} \sum_{\nu_{0} \in \{ 0,1,2, \dots , 7 \}}	\nonumber	\\
	&\times \left[\log \Gamma_B (T^{2}_{\nu_1 \nu_2 \nu_{3}}) - \log \Gamma_B (T^{2}_{\nu_0 \nu_2 \nu_{3}})
		+ \log \Gamma_B (T^{2}_{\nu_0 \nu_1 \nu_{3}}) - \log \Gamma_B (T^{2}_{\nu_0 \nu_1 \nu_2}) \right]	\nonumber	\\
	=& \frac{V^{(7)}}{8} \sum_{0 \leq r < \nu_1 < \nu_2 < \nu_3 \leq 7} (-1)^{r+\nu_1 + \nu_2 + \nu_3}
		\frac{1}{iV^{(4)}} \sqrt{\frac{5}{8}} \log \Gamma_{\Lambda}(T^{2, \hat{r} \hat{\nu}_1 \hat{\nu}_2 \hat{\nu}_3}_{012\dots 7})
		\frac{1}{iV^{(3)}} \sqrt{\frac{{4}}{{6}}}	\nonumber	\\
	&\times \sum_{\nu_{0} \in \{ 0,1,2, \dots , 7 \}}
		\left[ \left\{ \log \Gamma_B (T^{2}_{\nu_1 \nu_2 \nu_{3}}) - \log \Gamma_B (T^{2}_{\nu_0 \nu_2 \nu_{3}})
		+ \log \Gamma_B (T^{2}_{\nu_0 \nu_1 \nu_{3}}) - \log \Gamma_B (T^{2}_{\nu_0 \nu_1 \nu_2}) \right\} \right.	\nonumber	\\
	&- \left\{ \log \Gamma_B (T^{2}_{r \nu_2 \nu_{3}}) - \log \Gamma_B (T^{2}_{\nu_0 \nu_2 \nu_{3}})
		+ \log \Gamma_B (T^{2}_{\nu_0 r \nu_{3}}) - \log \Gamma_B (T^{2}_{\nu_0 r \nu_2}) \right\}	\nonumber	\\
	&+ \left\{ \log \Gamma_B (T^{2}_{r \nu_1 \nu_{3}}) - \log \Gamma_B (T^{2}_{\nu_0 \nu_1 \nu_{3}})
		+ \log \Gamma_B (T^{2}_{\nu_0 r \nu_{3}}) - \log \Gamma_B (T^{2}_{\nu_0 r \nu_1}) \right\}	\nonumber	\\
	&\left. - \left\{ \log \Gamma_B (T^{2}_{r \nu_1 \nu_{2}}) - \log \Gamma_B (T^{2}_{\nu_0 \nu_1 \nu_{2}})
		+ \log \Gamma_B (T^{2}_{\nu_0 r \nu_{2}}) - \log \Gamma_B (T^{2}_{\nu_0 r \nu_1}) \right\} \right]	\nonumber	\\
	=& \frac{V^{(7)}}{8} \sum_{0 \leq r < \nu_1 < \nu_2 < \nu_3 \leq 7} (-1)^{r+\nu_1 + \nu_2 + \nu_3}
		\frac{1}{iV^{(4)}} \sqrt{\frac{5}{8}} \log \Gamma_{\Lambda}(T^{2, \hat{r} \hat{\nu}_1 \hat{\nu}_2 \hat{\nu}_3}_{012\dots 7})
		\frac{1}{iV^{(3)}} \sqrt{\frac{{4}}{{6}}}	\nonumber	\\
	&\times \left[ 8\log \Gamma_B (T^{2}_{\nu_1 \nu_2 \nu_{3}}) - 8\log \Gamma_B (T^{2}_{r \nu_2 \nu_{3}})
	+ 8\log \Gamma_B (T^{2}_{r \nu_1 \nu_{3}}) - 8\log \Gamma_B (T^{2}_{r \nu_1 \nu_{2}}) \right]	\nonumber	\\
	=& \frac{V^{(6)} V^{(1)}}{7} \sqrt{\frac{8}{14}} \sum_{0 \leq r < \nu_1 < \nu_2 < \nu_3 \leq 7} (-1)^{r+\nu_1 + \nu_2 + \nu_3}
		\frac{1}{iV^{(4)}} \sqrt{\frac{5}{8}} \log \Gamma_{\Lambda}(T^{2, \hat{r} \hat{\nu}_1 \hat{\nu}_2 \hat{\nu}_3}_{012\dots 7})
		\frac{3}{iV^{(2)}V^{(1)}} \sqrt{\frac{6}{4}} \sqrt{\frac{{4}}{{6}}}	\nonumber	\\
	&\times \left[ \log \Gamma_B (T^{2}_{\nu_1 \nu_2 \nu_{3}}) - \log \Gamma_B (T^{2}_{r \nu_2 \nu_{3}})
	+ \log \Gamma_B (T^{2}_{r \nu_1 \nu_{3}}) - \log \Gamma_B (T^{2}_{r \nu_1 \nu_{2}}) \right]	\nonumber	\\
	=& \frac{V^{(6)}}{7} \sqrt{\frac{8}{14}} \sum_{0 \leq r < \nu_1 < \nu_2 < \nu_3 \leq 7} (-1)^{r+\nu_1 + \nu_2 + \nu_3}
		\frac{1}{iV^{(4)}} \sqrt{\frac{5}{8}} \log \Gamma_{\Lambda}(T^{2, \hat{r} \hat{\nu}_1 \hat{\nu}_2 \hat{\nu}_3}_{012\dots 7})
		\frac{3}{iV^{(2)}}	\nonumber	\\
	&\times \left[ \log \Gamma_B (T^{2}_{\nu_1 \nu_2 \nu_{3}}) - \log \Gamma_B (T^{2}_{r \nu_2 \nu_{3}})
	+ \log \Gamma_B (T^{2}_{r \nu_1 \nu_{3}}) - \log \Gamma_B (T^{2}_{r \nu_1 \nu_{2}}) \right].
	\label{D20}
\end{align}

Finally, redefining the label of vertices as $r \to q, \nu_1 \to r, \nu_2 \to \nu_1, \nu_3 \to \nu_2$, the right hand side of (\ref{D20}) becomes 
\begin{align}
	& \frac{V^{(6)}}{7} \sqrt{\frac{8}{14}} \sum_{0 \leq r < \nu_1 < \nu_2 < \nu_3 \leq 7} (-1)^{r+\nu_1 + \nu_2 + \nu_3}
		\frac{1}{iV^{(4)}} \sqrt{\frac{5}{8}} \log \Gamma_{\Lambda}(T^{2, \hat{r} \hat{\nu}_1 \hat{\nu}_2 \hat{\nu}_3}_{012\dots 7})
		\frac{3}{iV^{(2)}}	\nonumber	\\
	&\times \left[ \log \Gamma_B (T^{2}_{\nu_1 \nu_2 \nu_{3}}) - \log \Gamma_B (T^{2}_{r \nu_2 \nu_{3}})
	+ \log \Gamma_B (T^{2}_{r \nu_1 \nu_{3}}) - \log \Gamma_B (T^{2}_{r \nu_1 \nu_{2}}) \right]	\nonumber	\\
	=&\sqrt{\frac{8}{14}} \sum_{0 \leq q < r < \nu_1 < \nu_2 \leq 7} (-1)^{q+r+\nu_1 + \nu_2} \frac{V^{(6)}}{7} 
		\frac{1}{iV^{(4)}} \sqrt{\frac{5}{8}} \log \Gamma_{\Lambda}(T^{2, \hat{r} \hat{\nu}_1 \hat{\nu}_2 \hat{\nu}_3}_{012\dots 7})
		\frac{3}{iV^{(2)}}	\nonumber	\\
	&\times \left[ \log \Gamma_B (T^{2}_{r \nu_1 \nu_{2}}) - \log \Gamma_B (T^{2}_{q \nu_1 \nu_{2}})
	+ \log \Gamma_B (T^{2}_{q r \nu_{2}}) - \log \Gamma_B (T^{2}_{q r \nu_{1}}) \right].
\end{align}
This result agrees with \eqref{dLambdaWedgedB_LHS}, and we see that we can discretize both sides of \eqref{dLambdaWedgeB} on a simplex.

\end{document}